\documentclass[largeformat,suppldata]{interact}

\usepackage[utf8]{inputenc}
\usepackage[T1]{fontenc}
\usepackage{natbib}
\bibpunct[, ]{(}{)}{;}{a}{}{,}

\usepackage{amsmath,amssymb,bm}
\usepackage{siunitx}
\usepackage{graphicx}
\usepackage{booktabs}
\usepackage{tabularx}
\usepackage{longtable}
\usepackage{array}
\usepackage{multirow}
\usepackage{threeparttable}
\usepackage{xcolor}\usepackage{tikz}\usetikzlibrary{arrows.meta,positioning}
\usepackage{hyperref}
\usepackage[capitalise,noabbrev]{cleveref}
\usepackage{caption}
\usepackage{subcaption}
\usepackage{placeins}
\usepackage{float}
\usepackage{xspace}

\hypersetup{colorlinks=true,linkcolor=black,citecolor=black,urlcolor=black}
\newcolumntype{Y}{>{\raggedright\arraybackslash}X}
\newcolumntype{L}[1]{>{\raggedright\arraybackslash}p{#1}}
\newcommand{\kwhstep}{\si{\kilo\watt\hour}/step}

\newcommand{\field}[1]{\nolinkurl{#1}}
\newcommand{\citylearnvthree}{CityLearn v3\xspace}
\newcommand{\citylearnvtwo}{CityLearn v2\xspace}
\newcommand{\rec}{renewable energy community\xspace}

\articletype{RESEARCH ARTICLE}

\title{\citylearnvthree: A Configurable Simulation and Evaluation Framework for Realistic Control Studies of Renewable Energy Communities}

\author{
\name{Tiago Fonseca\textsuperscript{a}\thanks{CONTACT Tiago Fonseca. Email: \texttt{calof@isep.ipp.pt}}, Luis Lino Ferreira\textsuperscript{a}, Armando Sousa\textsuperscript{b}, Ava Mohammadi\textsuperscript{c} and Zoltan Nagy\textsuperscript{c}}
\affil{\textsuperscript{a}INESC TEC / Polytechnic of Porto - School of Engineering, Porto, Portugal; \textsuperscript{b}FEUP - Faculty of Engineering, University of Porto / INESC TEC, Porto, Portugal; \textsuperscript{c}Department of the Built Environment, Building Services, Eindhoven University of Technology, Eindhoven, the Netherlands}
}

\begin{document}
\maketitle

\begin{abstract}

Renewable energy communities (RECs) coordinate buildings, photovoltaic generation, batteries, electric vehicles and flexible loads. Controller studies often simplify changing participation, equipment availability, service deadlines and data quality, so lower cost or peak demand can conceal missed services or infeasible power requests. This paper presents CityLearn v3, a configurable simulation and evaluation framework for REC control studies under these conditions. It represents changing members and assets, flexible-load deadlines, demand-response requests, local energy sharing, and data or equipment failures within one simulation environment. Building and phase power limits constrain controllable requests, while a declared timestep preserves consistent power-to-energy accounting. The framework records controller inputs and distinguishes requested actions from those applied to the simulated equipment. Reference controllers, service- and constraint-aware performance indicators, and trajectory exports support comparisons within and across communities. Software checks and application examples examine service delivery, electrical constraints, settlement and changing scenarios; a synthetic high-frequency trace replay illustrates how aggregation can conceal short peaks without changing annual energy. Together, these records allow aggregate performance to be interpreted alongside service failures, action reductions and participant-level outcomes.
\end{abstract}


\section{Introduction}
\label{sec:introduction}

Buildings increasingly combine photovoltaic (PV) generation, batteries, heat pumps, flexible appliances and electric-vehicle (EV) chargers. Households, public buildings and businesses can coordinate these resources within a \rec to increase local PV use, reduce peak grid import, provide flexibility and share locally generated energy.

Consider a simple day as an example. In the morning, a new household joins the community and its PV system becomes part of the local energy balance. During the afternoon, an EV arrives at a charger and must reach a required state of charge before departure. A washing-machine or process cycle may start later, but it still has to finish before its deadline. In the evening, the grid operator may ask the community to reduce import for one hour. During the same period, a meter may stop reporting and one battery command may fail. A realistic controller study must state all these conditions and preserve them in the reported results.

Established district-control benchmarks provide shared scenarios and evaluation protocols \citep{citylearn_original,citylearnv2}. Studies using fixed membership and complete data can, however, leave the changing conditions described above outside the control problem. Established simulation tools cover different parts of the energy-system problem. EnergyPlus and the Modelica Buildings library provide detailed models of buildings, thermal systems and HVAC equipment; BOPTEST provides standardized test cases, an application programming interface and performance indicators for building-control benchmarking; and HELICS and mosaik connect models from different domains in a common co-simulation \citep{crawley2001_energyplus,wetter2014_modelica_buildings,blum2021_boptest,hardy2024_helics,steinbrink2019_mosaik}. CityLearn addresses a different need as it provides a common Gym-compatible environment in which control policies can be tested over the same energy scenarios, controller interfaces and performance indicators. The original CityLearn established this benchmark role \citep{citylearn_original}, and \citylearnvtwo extended it with rule-based control (RBC), model-predictive control (MPC), reinforcement-learning control (RLC), distributed energy resources, EV and vehicle-to-grid examples, occupant feedback and resilience studies \citep{citylearnv2,gymnasium,evlearn}. \Cref{tab:simulator_comparison} summarizes these complementary approaches in terms of their modelling scope, control-study interfaces, and representation of community services and constraints.\par{} \begin{table}[H]\centering\caption{Base capabilities of selected simulation and control-benchmarking platforms.}\label{tab:simulator_comparison}\footnotesize\setlength{\tabcolsep}{3pt}\renewcommand{\arraystretch}{1.15}\begin{tabularx}{\textwidth}{L{0.17\textwidth}YYY}\toprule Platform & Primary purpose & Control-study interface & Community and service scope \\\midrule{} EnergyPlus and Modelica Buildings \citep{crawley2001_energyplus,wetter2014_modelica_buildings} & Detailed building, HVAC and district-energy physics & Control interfaces and component models are available; the experiment protocol is assembled by the user & Community membership, local settlement and common controller-evaluation semantics are study-specific \\BOPTEST \citep{blum2021_boptest} & Repeatable benchmarking of building control using high-fidelity emulators & Standard API, test cases, baseline control and KPIs & Centred on building and HVAC control; REC services and community electrical constraints require additional modelling \\HELICS and mosaik \citep{hardy2024_helics,steinbrink2019_mosaik} & Synchronization and data exchange across heterogeneous simulators & Coordinates user-supplied models and controllers; evaluation logic is defined by the federation or scenario & Can support multi-domain REC studies, but REC, service and audit semantics are supplied by the coupled components \\\citylearnvtwo \citep{citylearnv2} & Data-driven district control benchmarking & Common Gym-compatible loop with reference and learning controllers and shared KPIs & Integrated buildings, distributed energy resources and EV studies within a shared district-control benchmark \\\bottomrule\end{tabularx}\end{table}\par{}

These approaches provide complementary foundations for modelling energy systems and benchmarking control. For the REC studies considered here, community membership, service obligations and electrical constraints must also remain consistent from scenario configuration through controller interaction to evaluation. To support such studies, this paper presents \citylearnvthree, a configurable simulation and evaluation framework informed by real-community projects. Its technical contribution is an integrated experimental interface that distinguishes requested from applied actions, enforces building, phase and equipment power limits, and preserves power-to-energy consistency across declared timesteps. Changing participants and assets, flexible services, local settlement and data or equipment failures can therefore be studied under a common control and evaluation protocol.

The distinction between requested and applied actions makes controller decisions auditable. A request to charge a disconnected EV remains recorded even though the applied charging power is zero; a battery request reduced by a phase limit is likewise retained alongside the applied value. Service indicators complement these records, so a reduction in cost can be assessed together with missed EV departures or flexible-load deadlines.

\Cref{sec:v2_to_requirements} derives the study requirements, and \cref{sec:environment} describes their representation. \Cref{sec:control_eval} sets out controller interfaces, reference policies, performance indicators, software checks and computational costs. \Cref{sec:applications} demonstrates their use in REC experiments, and \cref{sec:discussion} discusses the findings.

\par{}

\section{Requirements for realistic renewable energy community studies}
\label{sec:v2_to_requirements}

Experience developing and testing REC controllers in the OPEVA and DEMFLEX projects motivates two groups of requirements: the conditions a scenario must represent, and the information the simulator must expose for fair controller comparison.

\subsection{From real communities to simulator requirements}
\label{subsec:deployment_motivation}

Community composition, equipment, service obligations and sharing rules affect both the control problem and the interpretation of its results. A scenario must specify these assumptions together with its time resolution, electrical limits and data availability.

Recent REC studies point to the same need. Energy communities are socio-technical systems, so their performance depends on participant composition, technical design, local objectives and governance as well as total demand \citep{gjorgievski2021_energy_communities_review,eu2018_redii,eu2019_electricity_directive}. Modelling results change with participant types, technologies, demand profiles and energy-sharing rules \citep{belmar2023_modelling_recs}. Reviews of machine-learning applications likewise argue that data-driven control should reflect how communities are operated, not only abstract load-shifting objectives \citep{hernandez2022_ml_local_energy_communities}. Simulation studies combine buildings, PV, batteries, grid interaction, data availability and uncertainty \citep{motlagh2023_lec_design_simulation}; use 15-minute decisions, EV availability, trips, flexible-load windows, export limits, local exchange and forecast uncertainty \citep{friess2024_daily_operation_recs}; and show that allocation and settlement rules change how community benefits are distributed \citep{mello2024_energy_allocation_settlement}.

The same requirements appeared in the authors' use of CityLearn and EVLearn in EV and energy-community projects. The OPEVA/EnergAIze demonstrator (Horizon Europe grant 101097267) considered a REC with nine houses, an office building, a factory, more than 20 EV chargers, PV generation, batteries and different building loads \citep{softcps_energaize,opeva_project,opeva_cordis,evlearn}. Work towards control deployment in that setting made several assumptions concrete. Chargers, meters and PV systems produced data at different frequencies; values could be missing or noisy; commands could fail; measured PV and load traces required consistent energy accounting across timesteps; appliance and process cycles had start windows and completion deadlines; homes had contracted power and phase limits; local energy sharing changed participant costs; and REC members or equipment could enter or leave during the study.

DEMFLEX (COMPETE2030-FEDER-01657200-19215) added the need to compare controllers across communities, represent demand-response requests explicitly, provide structured inputs to advanced controllers and export trajectories and performance indicators that can be inspected outside the simulator \citep{demflex_softcps,demflex_eurogia}. These projects motivate the requirements below.

\subsection{Scenario and evaluation requirements}
\label{subsec:requirements}
\label{subsec:requirements_to_design}

\Cref{tab:requirements} summarizes the scenario requirements, developed in \cref{sec:environment}.

{\footnotesize
\begin{longtable}{p{0.08\linewidth}p{0.25\linewidth}p{0.29\linewidth}p{0.26\linewidth}}
\caption{Requirements for realistic REC scenarios.}\label{tab:requirements}\\
\toprule
ID & REC requirement & Real example & How \citylearnvthree represents it \\
\midrule
\endfirsthead
\toprule
ID & REC requirement & Real example & How \citylearnvthree represents it \\
\midrule
\endhead
D1 & Multiple communities & A project may study several communities but must retain the result of each one. & Synchronized community simulations with local and combined results. \\
D2 & Demand-response requests & A grid operator requests a power change during a stated time window. & Time-stamped interval DR specifications, delivered response, shortfall and payment. \\
D3 & Local energy sharing and settlement & PV surplus can meet another member's demand and change both members' costs. & Participant-level local import, local export, and settlement with the grid and community. \\
D4 & Changing community members and assets & A household joins, a charger is installed or a PV system is removed during the study. & Member and equipment changes that update controller inputs, actions and results. \\
D5 & Flexible-load schedules and deadlines & An appliance or process may start operation later, but must complete before a deadline. & Cycle profiles, allowed start windows, deadlines and service indicators. \\
D6 & Data and equipment failures & A meter value is missing, a forecast is biased, a command is lost or a battery is unavailable. & Repeatable changes to measurements, forecasts, commands and equipment availability. \\
D7 & Time resolution and energy units & Data may arrive every 15 s, 15 min or hour, while power and energy must remain distinct. & A declared timestep and conversion checks. \\
D8 & Building and phase power limits & A home has contracted import/export limits and equipment connected to specific phases. & Building-level and per-phase limits applied to controllable power requests. \\
\bottomrule
\end{longtable}
}

\Cref{tab:simulator_quality_requirements} summarizes the evaluation requirements: controller inputs and actions, performance indicators, reference policies, software validation and computational cost. \Cref{sec:control_eval} describes how these are addressed.

{\footnotesize
\begin{longtable}{p{0.08\linewidth}p{0.25\linewidth}p{0.29\linewidth}p{0.26\linewidth}}
\caption{Requirements for fair and reproducible controller comparisons.}\label{tab:simulator_quality_requirements}\\
\toprule
ID & Evaluation requirement & Why it matters & How \citylearnvthree addresses it \\
\midrule
\endfirsthead
\toprule
ID & Evaluation requirement & Why it matters & How \citylearnvthree addresses it \\
\midrule
\endhead
S1 & Controller inputs and applied actions & The user must know what the controller received, requested and actually caused. & Structured observations with persistent identities, availability masks, feasible action ranges and separate requested/applied actions. \\
S2 & Performance indicators and result exports & One reward or aggregate curve can hide service failures and local outcomes. & Extended KPIs with energy, cost, emissions, service and constraint indicators with time-series and comparison exports. \\
S3 & Reference controllers and fair comparisons & A normalized result is meaningful only when the comparison case uses the same given configuration and accounting rules. & Business-as-usual (BAU) and rule-based reference controllers evaluated in the same scenario. \\
S4 & Software validation & More scenario rules create more opportunities for plausible but incorrect results. & Regression tests, physical checks, interface checks, indicator-consistency checks and saved evidence. \\
S5 & Computational cost & High-frequency data, structured inputs and several communities can slow repeated controller studies. & Measurements of run time, memory usage, data export time, data format. \\
\bottomrule
\end{longtable}
}

\section{Representing realistic REC scenarios in CityLearn v3}
\label{sec:environment}

The scenario represents the eight REC requirements in \cref{tab:requirements}. The following subsections describe their physical or service meaning, configuration and recorded outcomes.

\begin{figure}[p] \centering \includegraphics[width=\linewidth,height=0.91\textheight,keepaspectratio]{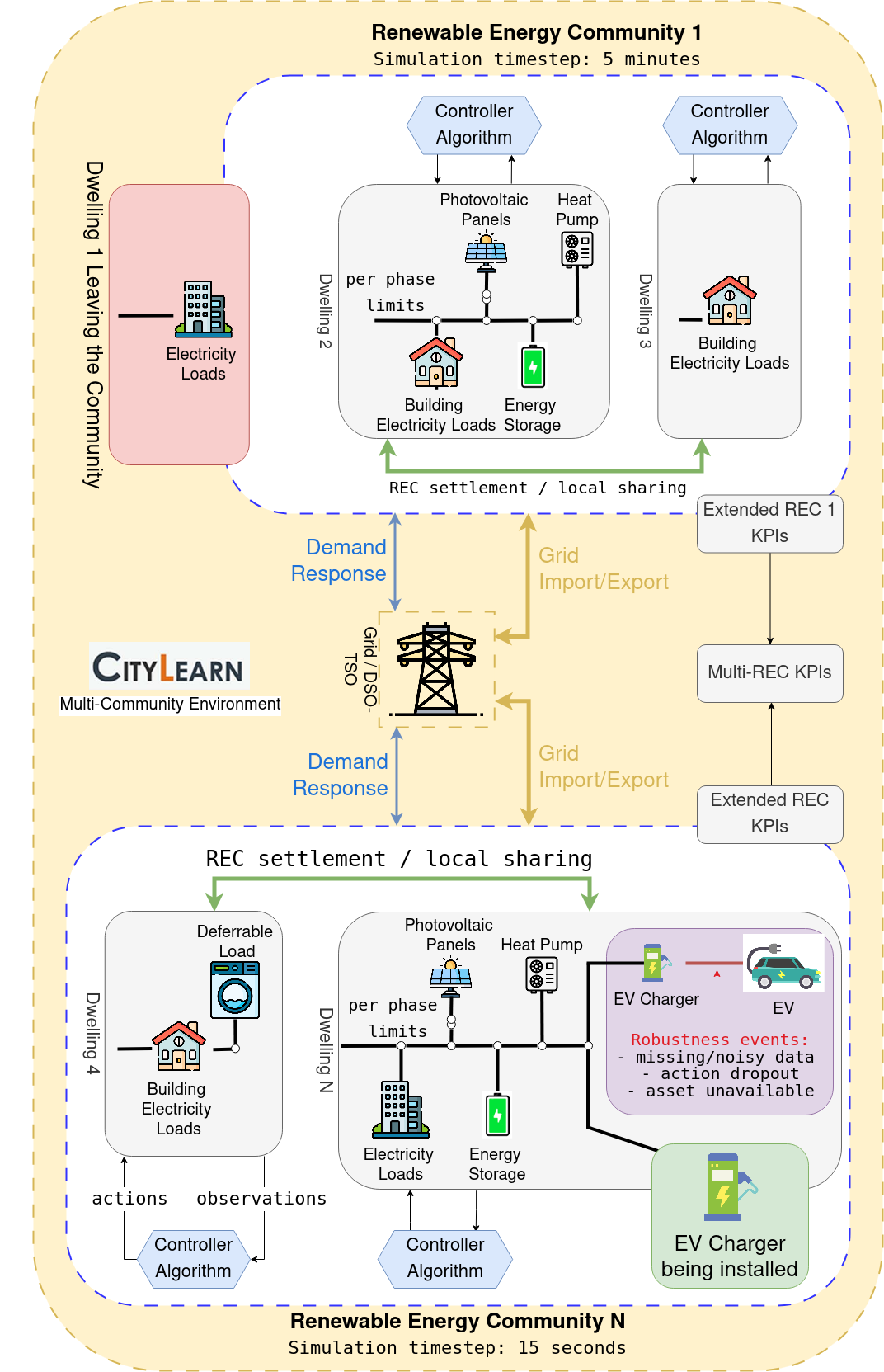} \caption{How CityLearn v3 represents one or more renewable energy communities.} \label{fig:environment_architecture} \end{figure}

\Cref{fig:environment_architecture} depicts the main simulator modules. Each blue dashed area is one REC; the yellow area shows that several RECs can be run together without losing their individual results. Within each REC, members exchange locally generated energy before the remaining import or export is accounted for at the grid. The arrows from the Grid/DSO-TSO represent demand-response requests, and the hexagonal controller boxes show the information and actions exchanged at each simulation step.

The examples inside the RECs illustrate membership changes, charger installation, flexible-load deadlines and per-phase limits. The box labelled \emph{robustness events} represents missing data, lost commands and unavailable equipment. Appendix~\ref{app:datasets}, \cref{tab:schema_config_map}, lists the corresponding configuration fields.

\subsection{Multiple communities}
\label{subsec:multi_community}

Requirement D1 is the ability to study several communities in one synchronized experiment. A municipality or aggregator may coordinate RECs with different buildings, resources and objectives. A policy can receive observations and provide actions for all communities at the same timestep, while their separate identities and results reveal which community caused a peak, produced surplus or missed a service. The wrapper synchronizes electrically independent community simulations and retains their individual and combined results.

\citylearnvthree runs community environments side by side. Each retains its scenario, members, equipment, demand-response requests, settlement rules and results. The wrapper resets them together and advances each once per simulation step. It accepts actions organized by community identifier and returns observations, rewards and information under those identifiers. The D1 row of \cref{tab:schema_config_map} lists the configuration fields.

The communities evaluated together must use the same centralized or decentralized setting, timestep and episode length. Individual scenarios can use other time resolutions when evaluated separately, but they must be aligned to the same simulation clock before they are included in one multi-community experiment.

Results are reported for each community and for the portfolio. Energy, cost, emissions, payments and event counts are summed; ratios and percentages use declared reporting weights, equal by default. Thus, self-consumption values of 40\% and 60\% give a mean of 50\%, or 55\% with weights of 1 and 3. Weights can reflect floor area, member count or annual demand and define averages of community-level ratios.

\subsection{Demand-response requests}
\label{subsec:demand_response}

Requirement D2 concerns explicit demand-response requests \citep{rl_demand_response_review}. A distribution system operator (DSO), transmission system operator (TSO), retailer or aggregator may ask a community to change its net consumption during a stated interval. The request and its delivery must be visible to the controller and in the results, rather than represented only through a tariff or reward.

Each request specifies its identifier, issuer, start and end timesteps, direction $q_r$, target power $R_r$, tolerance $\epsilon_r$, payment rate and shortfall penalty. Delivery is measured against a reference baseline $B_{r,t}$: mean community net power over a configurable pre-event window, one hour by default, calculated immediately before activation and held fixed during the event. This pre-event reference is used consistently for response measurement and settlement. Invalid or incomplete history invalidates the baseline and excludes the request from settlement.

With positive $P_t$ denoting net import, a \emph{down} request asks the community to reduce import relative to the baseline, whereas an \emph{up} request asks it to increase net consumption by increasing import or reducing export. For an active request $r$, with direction $q_r$ and target power $R_r$, the delivered power is
\begin{equation}
d_{r,t} =
\begin{cases}
P_t - B_{r,t}, & q_r=\mathrm{up},\\
B_{r,t} - P_t, & q_r=\mathrm{down},
\end{cases}
\label{eq:dr_delivery}
\end{equation}
Here $P_t$ is actual community net power, positive for import. A positive $d_{r,t}$ indicates movement in the requested direction; a negative value indicates the opposite.

The simulator derives credited power $c_{r,t}$ and shortfall power $s_{r,t}$ from the delivered power. Credited power is the non-negative response that counts towards payment and is capped at the target $R_r$. Shortfall is the part still missing after the tolerance $\epsilon_r$:
\begin{equation}
    c_{r,t} = \min(\max(d_{r,t},0), R_r), \qquad
    s_{r,t} = \max(R_r - d_{r,t} - \epsilon_r, 0),
    \label{eq:dr_credit_shortfall}
\end{equation}
Power is converted to energy by multiplying by the timestep duration in hours. For example, a \SI{20}{\kilo\watt} down request during one \SI{15}{\minute} step asks for \SI{5}{\kilo\watt\hour}. If import falls by only \SI{14}{\kilo\watt} and the tolerance is \SI{1}{\kilo\watt}, the credited response is \SI{3.5}{\kilo\watt\hour} and the shortfall is \SI{1.25}{\kilo\watt\hour}. If import falls by \SI{23}{\kilo\watt}, the full delivery remains visible, but credit is capped at the \SI{20}{\kilo\watt} target and shortfall is zero.

A scenario is defined by a top-level configuration file, normally \texttt{schema.json}. Its \texttt{demand\_response} block enables the service, selects the baseline window and points to a CSV or Parquet request file containing one row per request. The current schema accepts DSO or TSO issuers and non-overlapping requests. During an active request, the structured input includes its identifier, issuer, direction, target power and baseline information. Appendix~\ref{app:datasets}, \cref{tab:schema_config_map}, gives the exact fields.

\subsection{Local energy sharing and settlement}
\label{subsec:rec_accounting}

Requirement D3 concerns local energy sharing and settlement. In European REC and citizen-energy-community policy, participation involves collective activity and member or local benefit, not only independent behind-the-meter operation \citep{eu2018_redii,eu2019_electricity_directive}. A member with PV surplus may allocate part of that energy to a neighbour before the remaining export is accounted for at the grid. Batteries, EV charging and flexible loads can then be scheduled to coincide with local surplus. The simulation must retain both the physical allocation and its effect on each member's cost.

District net load cannot identify who supplied or consumed shared energy, or how savings were distributed. For example, simultaneous office PV export and household import can cancel in the aggregate whether or not local sharing is enabled. \citylearnvthree therefore records participant net consumption, local exchange, settlement and residual grid import/export separately.

Let $E^{\mathrm{net}}_{b,t}$ be the net energy of building $b$ during timestep $t$ after local generation and controllable equipment. Positive values are demand and negative values are surplus. For the active members $\mathcal{C}_t$, total demand $D_{\mathcal{C},t}$ and total surplus $S_{\mathcal{C},t}$ are
\begin{align}
    D_{\mathcal{C},t} = \sum_{b \in \mathcal{C}_t}\max(E^{\mathrm{net}}_{b,t},0), \qquad
    S_{\mathcal{C},t} = \sum_{b \in \mathcal{C}_t}\max(-E^{\mathrm{net}}_{b,t},0).
    \label{eq:community_demand_surplus}
\end{align}
When local sharing is enabled, the shared energy is the smaller of demand and surplus:
\begin{equation}
    L_{\mathcal{C},t} = \min(D_{\mathcal{C},t}, S_{\mathcal{C},t});
    \label{eq:local_traded_energy}
\end{equation}
if sharing is disabled, $L_{\mathcal{C},t}=0$. Participant weights decide how $L_{\mathcal{C},t}$ is distributed among the members that are importing energy  without exceeding their demand. Exporting members contribute in proportion to their surplus. Residual grid import $G^+_{\mathcal{C},t}$ is total demand minus shared energy, and residual grid export $G^-_{\mathcal{C},t}$ is total surplus minus shared energy. The allocation and price settings determine the resulting settlement.
As a simple example, suppose that one member has \SI{10}{\kilo\watt\hour} of PV surplus during a timestep and two other members together demand \SI{6}{\kilo\watt\hour}. If all members are eligible for local sharing, the community layer records \SI{6}{\kilo\watt\hour} of local traded energy and \SI{4}{\kilo\watt\hour} of residual export. If a controller shifts an additional appliance or EV charging session into that same period, local traded energy can increase and residual grid exchange can fall. Participant-level records show who supplied and consumed the shared energy and how the configured rule affected costs.

Settlement is calculated from this local energy record rather than inferred later from aggregate net load. The member-level local price $p_{b,t}^{\mathrm{loc}}$ is the retail import price $p_{b,t}^{\mathrm{buy}}$ multiplied by the configured ratio $\rho$, where $0\leq\rho\leq1$. A value below one prices locally shared energy below retail import. For local import $L^{\mathrm{in}}_{b,t}$, local export $L^{\mathrm{out}}_{b,t}$ and residual grid import $G^{+}_{b,t}$, the current settlement cost is
\begin{equation}
    C_{b,t} =
    p_{b,t}^{\mathrm{buy}}G^{+}_{b,t}
    + p_{b,t}^{\mathrm{loc}}L^{\mathrm{in}}_{b,t}
    - p_{b,t}^{\mathrm{loc}}L^{\mathrm{out}}_{b,t}.
    \label{eq:participant_settlement}
\end{equation}
Imports are costs and local exports are credits. Residual grid export $G^{-}_{b,t}$ remains in the physical record but has zero remuneration in the current implementation, so it does not appear in this cost equation. Total local import and local export are both equal to the shared energy in \cref{eq:local_traded_energy}. The monetary record is budget-balanced when matched energy is charged and credited at one common local price. With different retail tariffs, local credits and charges follow each member's derived local price; a common local price gives exact bilateral balance.

Continuing the example above, if the retail import price is 0.25 EUR/kWh and $\rho=0.8$, local energy is settled at 0.20 EUR/kWh. A member importing \SI{6}{\kilo\watt\hour} locally pays 1.20 EUR instead of 1.50 EUR from the grid, while the producing member receives a local credit for the energy used by the community. The exported evidence therefore contains both the physical energy allocation and the economic effect: local import, local export, residual grid import/export, settled cost, counterfactual cost, local-market savings and participant/community indicators.

The \texttt{community\_market} block enables sharing, sets the local-price ratio and optionally supplies import-member weights. These weights allocate scarce surplus among importing members; omitted weights give equal default eligibility. The D3 row of \cref{tab:schema_config_map} lists the fields, exported indicators and legacy price-ratio alias.

\subsection{Changing community members and assets}
\label{subsec:dynamic_topology}

Requirement D4 concerns members and equipment that enter or leave during a study. A household may join, a charger may be installed, or a PV system may be removed for maintenance. Representing these changes within one run lets the controller experience the transition while preserving consistent inputs, actions and accounting across periods.

\citylearnvthree stores each change as a time-stamped record with an operation (add or remove), the affected member or asset identifier and, when needed, replacement parameters such as input files, installed power or phase connection. All changes scheduled for timestep $t$ are applied before the controller acts at that timestep. Demand, generation, available actions, power limits, local settlement and service checks therefore use the same active member and asset lists.

\par{}

In the scenario file, this behaviour is activated with \texttt{topology\_mode="dynamic"} and a list of \texttt{topology\_events}. The complete schema stores the events as validated records, but conceptually the declaration is close to:
\begin{quote}
\small
\texttt{t=1000: add member M4 with demand and PV traces}\\
\texttt{t=1300: add EV charger to M2}\\
\texttt{t=1600: remove PV from M1}\\
\texttt{t=2000: remove member M3}
\end{quote}
Add records can reuse an existing member or item of equipment as a template and replace selected parameters. Remove records close the corresponding active period. These changes are declared explicitly rather than hidden as zero-filled time series or undocumented preprocessing.

Performance indicators follow the active period of each member and asset. A charger installed at $t=1300$, for example, has no service denominator or action before that timestep. A removed member keeps its historical rows but does not contribute to later settlement. The exported change log explains why member counts, service denominators, local sharing or power-limit results changed during the episode.

\subsection{Flexible-load schedules and deadlines}
\label{subsec:deferrables}

Requirement D5 concerns flexible loads that must deliver a complete service, such as a dishwasher program, water-heating cycle, pump schedule or industrial batch. The controller chooses when the predefined cycle starts, and an omitted cycle is recorded as an unserved service.

In the implementation, these are configured as \texttt{deferrable\_appliances} using two linked files. The cycle file gives the energy profile $q_{c,\tau}$ in \kwhstep{}, its duration $d_c$ in integer timesteps and total energy. The schedule file gives the earliest start $t_c^{\mathrm{earliest}}$, latest start $t_c^{\mathrm{latest}}$, completion deadline $t_c^{\mathrm{deadline}}$, priority and whether the cycle is mandatory. The controller selects the start time of the stored energy profile.

For a requested cycle $c$, an applied start $s_c$ is feasible only if
\begin{equation}
    s_c \in \mathbb{Z}, \qquad
    t_c^{\mathrm{earliest}} \leq s_c \leq t_c^{\mathrm{latest}}, \qquad
    s_c + d_c - 1 \leq t_c^{\mathrm{deadline}},
    \label{eq:deferrable_start_feasibility}
\end{equation}
where $s_c$ is the selected start timestep. The term $s_c+d_c-1$ is the last timestep occupied by a $d_c$-step cycle, so the final inequality ensures completion by the inclusive deadline. The start must also be compatible with the configured equipment and building limits. Once started, the cycle contributes
\begin{equation}
    E^{\mathrm{def}}_{c,t} =
    \begin{cases}
    q_{c,t-s_c}, & s_c \leq t < s_c+d_c,\\
    0, & \mathrm{otherwise}.
    \end{cases}
    \label{eq:deferrable_profile}
\end{equation}
Here $E^{\mathrm{def}}_{c,t}$ is the cycle energy at timestep $t$ and $q_{c,t-s_c}$ selects the corresponding element of the stored profile. Completed cycles, missed cycles, start delay, served energy and unserved energy are then exported as performance indicators.

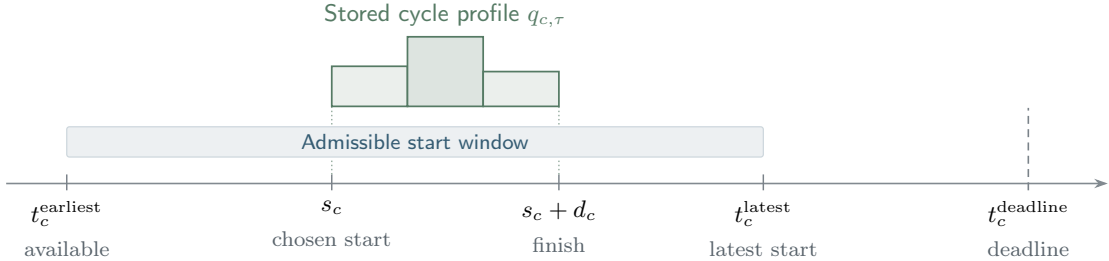
\begin{figure}[H]
    \centering
    \definecolor{cycleblue}{HTML}{355C73} \definecolor{cyclegreen}{HTML}{557A64} \definecolor{cyclegray}{HTML}{59636B} \begin{tikzpicture}[x=1cm,y=1cm, every node/.style={font=\sffamily}, event/.style={align=center,inner sep=0pt,font=\fontsize{9}{11}\selectfont}, panel/.style={draw=cycleblue!30,fill=cycleblue!3,rounded corners=2pt,line width=0.55pt,text width=4cm,minimum height=1.15cm,inner sep=5pt,align=left,font=\fontsize{9}{11}\selectfont\sffamily}, link/.style={-{Stealth[length=1.7mm,width=1.25mm]},draw=cycleblue!65,line width=0.65pt,shorten <=2mm,shorten >=2mm}] \path[fill=cycleblue!9,draw=cycleblue!25,line width=0.45pt,rounded corners=1pt] (0.95,0.35) rectangle (10.15,0.75); \node[text=cycleblue,font=\fontsize{8.5}{10}\selectfont\sffamily] at (5.55,0.55) {Admissible start window}; \draw[draw=cyclegray!75,line width=0.6pt,-{Stealth[length=1.8mm,width=1.25mm]}] (0.15,0) -- (14.7,0); \foreach \x in {0.95,4.45,7.45,10.15} {\draw[cyclegray!80,line width=0.55pt] (\x,-0.09) -- (\x,0.12);} \draw[draw=cyclegray!80,densely dashed,line width=0.6pt] (13.65,-0.09) -- (13.65,1.0); \foreach \x in {4.45,7.45} {\draw[cyclegreen!60,densely dotted,line width=0.55pt] (\x,0.12) -- (\x,0.32); \draw[cyclegreen!60,densely dotted,line width=0.55pt] (\x,0.78) -- (\x,1.02);} \path[fill=cyclegreen!12,draw=cyclegreen,line width=0.65pt] (4.45,1.02) rectangle (5.45,1.55); \path[fill=cyclegreen!20,draw=cyclegreen,line width=0.65pt] (5.45,1.02) rectangle (6.45,1.94); \path[fill=cyclegreen!12,draw=cyclegreen,line width=0.65pt] (6.45,1.02) rectangle (7.45,1.48); \node[text=cyclegreen,font=\fontsize{9}{11}\selectfont\sffamily] at (5.95,2.22) {Stored cycle profile $q_{c,\tau}$}; \node[event,anchor=north] at (0.95,-0.22) {$t_c^{\mathrm{earliest}}$\\[2pt]\textcolor{cyclegray}{\fontsize{8}{10}\selectfont available}}; \node[event,anchor=north] at (4.45,-0.22) {$s_c$\\[2pt]\textcolor{cyclegray}{\fontsize{8}{10}\selectfont chosen start}}; \node[event,anchor=north] at (7.45,-0.22) {$s_c+d_c$\\[2pt]\textcolor{cyclegray}{\fontsize{8}{10}\selectfont finish}}; \node[event,anchor=north] at (10.15,-0.22) {$t_c^{\mathrm{latest}}$\\[2pt]\textcolor{cyclegray}{\fontsize{8}{10}\selectfont latest start}}; \node[event,anchor=north] at (13.65,-0.22) {$t_c^{\mathrm{deadline}}$\\[2pt]\textcolor{cyclegray}{\fontsize{8}{10}\selectfont deadline}};  \end{tikzpicture}
    \caption{Flexible-load start window, deadline and energy cycle.}
    \label{fig:deferrable_cycle}
\end{figure}

\Cref{fig:deferrable_cycle} relates the admissible start window to the stored cycle profile and completion deadline. The resulting service record identifies completed and missed cycles, start delays, and served or unserved energy.

\subsection{Data and equipment failures}
\label{subsec:robustness_events}

Requirement D6 concerns failures in the information and equipment used by a controller. \citylearnvthree distinguishes four types of data and equipment failures:

\begin{description}
    \item[Measurement failures.] These change the current values received from meters or sensors. A building-load measurement can, for example, become missing, noisy, biased, stuck at an earlier value or clipped to a specified range. The physical load remains unchanged; only the information received by the controller is affected.
    
    \item[Forecast failures.] These change predicted values used by the controller, such as future PV generation, electricity prices or building demand. For example, a price forecast can be shifted by a fixed bias while the actual electricity price remains unchanged.
    
    \item[Action failures.] These affect the command sent by the controller before it reaches the simulated equipment. A battery-charging command can, for example, be lost, delayed, biased, noisy, stuck or clipped. The requested action remains recorded separately from the action that is actually applied.
    
    \item[Equipment failures.] These make an item of equipment temporarily unavailable for measurement, control or both. For example, a battery or EV charger can remain physically present in the scenario while rejecting control actions during a declared failure interval.
\end{description}

\citylearnvthree stores these cases as time-stamped data and equipment failures, configured internally through the \texttt{robustness} block present in the schema. Each record states the affected channel, target, feature, start and end timesteps, failure mode and parameters.

The four failure types enter the control loop at different points. Let $\mathcal{E}^{o}_t$, $\mathcal{E}^{f}_t$, $\mathcal{E}^{a}_t$ and $\mathcal{E}^{u}_t$ be the active measurement, forecast, action and equipment-availability failures at timestep $t$. Their effect is
\begin{align}
    \tilde{o}_t &= \rho_o(o_t,\mathcal{E}^{o}_t,\mathcal{E}^{u}_t), &
    \tilde{f}_t &= \rho_f(f_t,\mathcal{E}^{f}_t), \nonumber\\
    a_t^{\mathrm{app}} &= \rho_a(a_t^{\mathrm{req}},\mathcal{E}^{a}_t,\mathcal{E}^{u}_t), &
    x_{t+1} &= F(x_t,a_t^{\mathrm{app}},w_t).
    \label{eq:robustness_channels}
\end{align}
Here $o_t$ and $f_t$ are unchanged inputs, and $\tilde{o}_t$ and $\tilde{f}_t$ are the versions received by the controller. The maps $\rho_o$ and $\rho_f$ apply measurement and forecast failures. For compactness, $\rho_a$ denotes command processing through failures, availability and feasibility checks, yielding $a_t^{\mathrm{app}}$ from $a_t^{\mathrm{req}}$. The state-update function $F$ advances physical state $x_t$ using that applied action and exogenous inputs $w_t$, such as demand, weather and PV generation. A missing measurement therefore does not become a physical load reduction, and a lost command remains distinguishable from the requested action.

\begin{figure}[H]
    \centering
    \includegraphics[width=0.98\linewidth]{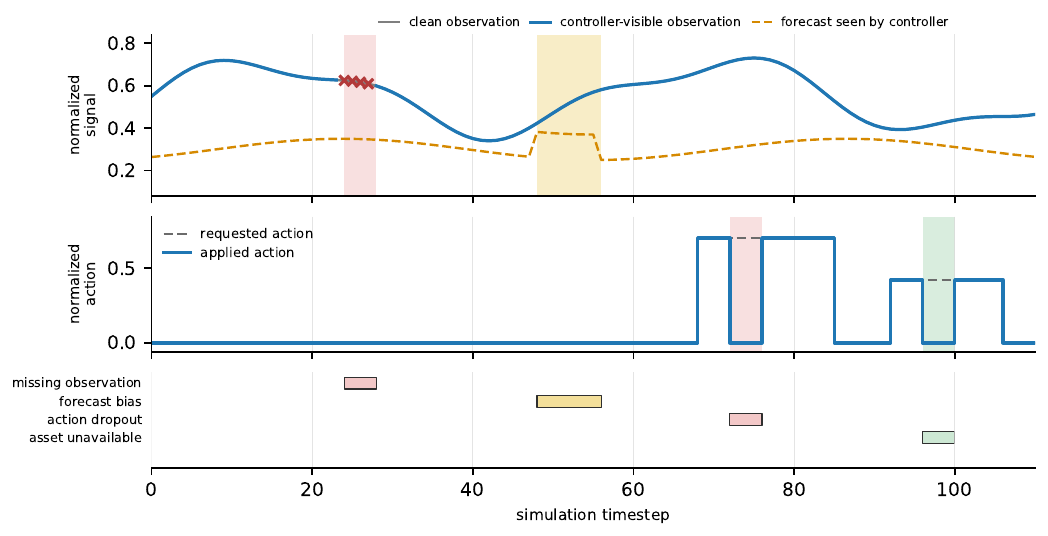}
    \caption{Timeline of data and equipment failures in the controller loop.}
    \label{fig:robustness_event_timeline}
\end{figure}

A short scenario can therefore describe common field problems without changing the controller implementation. The example used here declares a missing building-load measurement at timesteps 24--27, a biased price forecast at 48--55, lost storage commands at 72--75 and unavailable storage at 96--99. A fixed random seed makes stochastic changes repeatable, and a configured marker distinguishes a missing measurement from a physical zero. \Cref{fig:robustness_event_timeline} shows the same example as a timeline: measurement and forecast failures change what the controller receives, while action and equipment failures change what the simulator applies. The D6 row of \cref{tab:schema_config_map} in Appendix~\ref{app:datasets} gives the exact fields of the internal \texttt{robustness} block and its event file.

The results report failure counts, active timesteps, changed measurements and forecasts, lost commands and equipment-unavailable timesteps. These records link controller behaviour to the conditions under which it was tested.

\subsection{Time resolution and energy units}
\label{subsec:unit_contract_detail}

Requirement D7 makes the simulation clock and energy units explicit. The timestep determines which variations the controller observes, when requests and deadlines occur, and how power is converted to energy. Hourly averages can hide short EV peaks, missing-data intervals or phase-limit exceedances. \citylearnvthree uses the configured timestep consistently across scenario timing, equipment accounting and evaluation; examples in this paper include hourly, 15 min, one-minute and 15 s data.

\citylearnvthree makes this clock explicit through the \texttt{seconds\_per\_time\_step} property in the scenario schema and the declared simulation and episode bounds. Once loaded, request windows, service deadlines, actions, rewards and result exports use this timestep. If timestamps are available, the loader records their inferred spacing so that 15 s data are not silently interpreted as 15 min or 1 h data. The D7 row of \cref{tab:schema_config_map} lists the exact clock and simulation-window fields.

The key conversion is simple. Let $\Delta t$ be the simulation step duration in seconds. If a measured signal is provided as average power $P_t$ in \si{\kilo\watt}, the energy associated with one simulation step is
\begin{equation}
    E_t = P_t \frac{\Delta t}{3600},
    \label{eq:power_energy_conversion}
\end{equation}
where $E_t$ is energy in \kwhstep{}, $P_t$ is average power in \si{\kilo\watt} and $\Delta t/3600$ converts seconds to hours. Load, PV and charging time series expressed as energy are read directly in \kwhstep{}; equipment ratings and import/export limits remain in \si{\kilo\watt}; and prices and carbon intensities remain per \si{\kilo\watt\hour}. Absolute measured series are distinguished from normalized profiles that must be scaled by capacity. For example, a \SI{7.2}{\kilo\watt} charger operating for one step represents \SI{7.2}{\kilo\watt\hour} at 1 h resolution, \SI{1.8}{\kilo\watt\hour} at 15 min and \SI{0.03}{\kilo\watt\hour} at 15 s.

\subsection{Building and phase power limits}
\label{subsec:electrical_service}

Requirement D8 concerns the power available at a building connection. A house, office or factory can have a contracted limit on the total power imported from or exported to the grid. A three-phase connection can also have a separate limit on each phase. Both types of limits must be respected at the same time. For example, a building can remain below its total import limit while exceeding the limit of the phase to which an EV charger is connected.

Phase assignment limits the power available to each device: a charger connected to L1 cannot use spare capacity on L2 or L3, whereas a balanced three-phase battery distributes its power across all three phases.

Consider a house with a total import limit of \SI{12}{\kilo\watt} and phase limits of \SI{7}{\kilo\watt}, \SI{5}{\kilo\watt} and \SI{4}{\kilo\watt} for L1, L2 and L3, respectively. Before any new control action, the house imports \SI{4}{\kilo\watt} on L1, \SI{2}{\kilo\watt} on L2 and \SI{1}{\kilo\watt} on L3, giving a total import of \SI{7}{\kilo\watt}. An EV charger connected to L1 then requests \SI{7.2}{\kilo\watt}. Accepting the complete request would increase total import to \SI{14.2}{\kilo\watt} and L1 import to \SI{11.2}{\kilo\watt}. The total connection has \SI{5}{\kilo\watt} of remaining capacity, but L1 has only \SI{3}{\kilo\watt}. The phase limit is therefore the restrictive one: the simulator applies \SI{3}{\kilo\watt} of charging and records that \SI{4.2}{\kilo\watt} of the request was rejected.

The same calculation is applied generally to all controllable electrical equipment. For building $b$, let $\Phi_b$ be its phase set: $\{\mathrm{L1}\}$ for a single-phase connection or $\{\mathrm{L1},\mathrm{L2},\mathrm{L3}\}$ for a three-phase connection. The set $\mathcal{U}_{b,t}$ contains the controllable equipment available at timestep $t$, such as EV chargers and stationary batteries. A request $u_{a,t}$ from equipment $a$ is signed power in \si{\kilo\watt}, with positive values increasing import and negative values increasing export. The fraction $m_{a,\phi}$ assigns this request to phase $\phi$ and sums to one across the phases used by the equipment. For example, $m_{a,\mathrm{L1}}=1$ for a charger connected only to L1, while a balanced three-phase battery assigns one third of its power to each phase.

Before applying the limits, the requested total and phase powers are
\begin{align}
    P^{\mathrm{req}}_{b,t}
    &=
    P^{0}_{b,t}
    +
    \sum_{a\in\mathcal{U}_{b,t}}u_{a,t},\\
    P^{\mathrm{req}}_{b,\phi,t}
    &=
    P^{0}_{b,\phi,t}
    +
    \sum_{a\in\mathcal{U}_{b,t}}m_{a,\phi}u_{a,t},
    \qquad \phi\in\Phi_b .
    \label{eq:electrical_service_requested_power}
\end{align}
Here $P^{0}_{b,t}$ is building power before new controllable requests, and $P^{0}_{b,\phi,t}$ is its phase allocation. These terms include non-controllable demand, local generation and flexible-load cycles that have already started.

Let $I_b^{\max}$ and $X_b^{\max}$ denote the total import and export limits. Their phase equivalents are $I_{b,\phi}^{\max}$ and $X_{b,\phi}^{\max}$. With positive power denoting import and negative power denoting export, an applied power state is feasible when
\begin{equation}
\begin{aligned}
    -X_b^{\max}
    &\leq P_{b,t}
    \leq I_b^{\max},\\
    -X_{b,\phi}^{\max}
    &\leq P_{b,\phi,t}
    \leq I_{b,\phi}^{\max},
    \qquad \forall \phi\in\Phi_b .
\end{aligned}
\label{eq:electrical_service_feasible_region}
\end{equation}
An omitted limit is unbounded. Both total and phase constraints must hold; satisfying the total limit alone does not guarantee a feasible phase allocation.

When necessary, \citylearnvthree scales the controllable requests using factors $\alpha_{a,t}\in[0,1]$. The resulting applied powers are
\begin{align}
    P^{\mathrm{app}}_{b,t}
    &=
    P^{0}_{b,t}
    +
    \sum_{a\in\mathcal{U}_{b,t}}\alpha_{a,t}u_{a,t},\\
    P^{\mathrm{app}}_{b,\phi,t}
    &=
    P^{0}_{b,\phi,t}
    +
    \sum_{a\in\mathcal{U}_{b,t}}
    \alpha_{a,t}m_{a,\phi}u_{a,t}.
    \label{eq:electrical_service_applied_power}
\end{align}
A factor of one leaves a request unchanged, while a factor of zero rejects it completely. In the example, the charger receives $\alpha_{a,t}=3/7.2\approx0.417$. Its applied power is therefore \SI{3}{\kilo\watt}, resulting in \SI{10}{\kilo\watt} of total building import and exactly \SI{7}{\kilo\watt} on L1. The original \SI{7.2}{\kilo\watt} request remains recorded alongside the applied value.

When non-controllable building power exceeds a limit, the simulator applies the available controllable reductions and records any remaining exceedance, retaining the original non-controllable load.

The results retain requested and applied actions, total and phase power, remaining capacity and rejected controllable power. Requested-action pressure is reported separately from residual limit violations calculated from applied power histories. Power exceedances are converted to energy using the timestep duration. This distinction separates an infeasible controller request from a physical exceedance that remains after controllable requests have been reduced.

Each building's optional \texttt{electrical\_service} block declares single- or three-phase operation, total and per-phase import/export limits, default load allocation and equipment phase connections. Controllers may receive remaining total and phase capacity as inputs. When the block is absent, legacy CityLearn behaviour is retained unless older charger constraints are configured. Appendix~\ref{app:datasets}, \cref{tab:schema_config_map}, lists the fields.

\section{Running and evaluating controller studies}
\label{sec:control_eval}

Once the REC scenario is defined, a controller study must specify its inputs, reference policy, performance indicators and reporting procedure. This section addresses requirements S1--S5 in \cref{tab:simulator_quality_requirements}. \Cref{fig:experiment_workflow} summarizes the configuration choices from dataset and timestep to controller selection and result comparison. 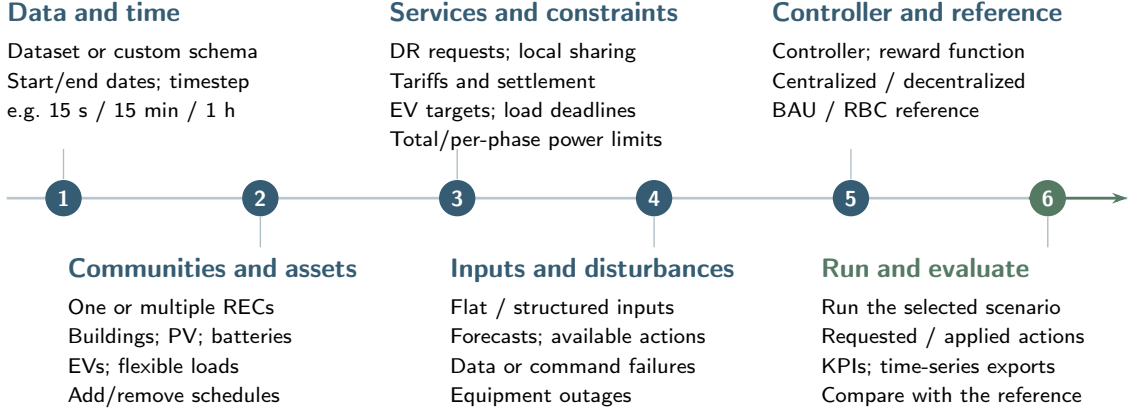
\begin{figure}[H] \centering \definecolor{workflowblue}{HTML}{355C73} \definecolor{workflowgreen}{HTML}{557A64} \begin{tikzpicture}[x=1cm,y=1cm, every node/.style={font=\sffamily}, choice/.style={text width=4.05cm,align=left,inner sep=0pt,font=\fontsize{8.5}{11}\selectfont\sffamily}, marker/.style={circle,fill=workflowblue,text=white,minimum size=4.8mm,inner sep=0pt,font=\fontsize{8}{8}\selectfont\sffamily\bfseries}, stem/.style={draw=workflowblue!40,line width=0.5pt}] \draw[workflowblue!35,line width=1pt] (0,0) -- (14.8,0); \draw[-{Stealth[length=2mm,width=1.4mm]},workflowgreen,line width=1pt] (13.5,0) -- (14.8,0); \node[marker] at (0.75,0) {1}; \node[marker] at (3.35,0) {2}; \node[marker] at (5.95,0) {3}; \node[marker] at (8.55,0) {4}; \node[marker] at (11.15,0) {5}; \node[marker,fill=workflowgreen] at (13.75,0) {6}; \draw[stem] (0.75,0.25) -- (0.75,0.65); \draw[stem] (3.35,-0.25) -- (3.35,-0.65); \draw[stem] (5.95,0.25) -- (5.95,0.65); \draw[stem] (8.55,-0.25) -- (8.55,-0.65); \draw[stem] (11.15,0.25) -- (11.15,0.65); \draw[stem,workflowgreen!50] (13.75,-0.25) -- (13.75,-0.65); \node[choice,anchor=north west] at (0,2.6) {\textcolor{workflowblue}{\bfseries\fontsize{10}{12}\selectfont Data and time}\\[3pt]Dataset or custom schema\\Start/end dates; timestep\\e.g.\ 15 s / 15 min / 1 h}; \node[choice,anchor=north west] at (0.8,-0.78) {\textcolor{workflowblue}{\bfseries\fontsize{10}{12}\selectfont Communities and assets}\\[3pt]One or multiple RECs\\Buildings; PV; batteries\\EVs; flexible loads\\Add/remove schedules}; \node[choice,anchor=north west] at (5.05,2.6) {\textcolor{workflowblue}{\bfseries\fontsize{10}{12}\selectfont Services and constraints}\\[3pt]DR requests; local sharing\\Tariffs and settlement\\EV targets; load deadlines\\Total/per-phase power limits}; \node[choice,anchor=north west] at (5.85,-0.78) {\textcolor{workflowblue}{\bfseries\fontsize{10}{12}\selectfont Inputs and disturbances}\\[3pt]Flat / structured inputs\\Forecasts; available actions\\Data or command failures\\Equipment outages}; \node[choice,anchor=north west] at (10.1,2.6) {\textcolor{workflowblue}{\bfseries\fontsize{10}{12}\selectfont Controller and reference}\\[3pt]Controller; reward function\\Centralized / decentralized\\BAU / RBC reference}; \node[choice,anchor=north west] at (10.75,-0.78) {\textcolor{workflowgreen}{\bfseries\fontsize{10}{12}\selectfont Run and evaluate}\\[3pt]Run the selected scenario\\Requested / applied actions\\KPIs; time-series exports\\Compare with the reference}; \end{tikzpicture} \caption{Main choices in a CityLearn v3 experiment. The selected dataset provides initial settings; the user configures assets, services, constraints and control before running the study and comparing its results.} \label{fig:experiment_workflow} \end{figure}

\subsection{Controller inputs and applied actions}
\label{subsec:control_configurations}
\label{subsec:entity_interface}
\label{subsec:action_masks}
\label{subsec:runtime_semantics}

Requirement S1 concerns the meaning of controller inputs and actions throughout a run. When membership, equipment or service availability changes, a vector position alone may no longer identify the same entity. The interface must retain identity, availability and feasible actions.

\citylearnvthree keeps the original CityLearn loop: a controller resets the environment, receives inputs, submits actions through \texttt{step}, and receives a reward and run information. The scenario selects the input format and action-feedback settings together with the timestep, services, failures, reference controller, reward, performance indicators and exports.

Two input formats are supported. The flat format preserves the CityLearn/Gymnasium vector used by existing RBC, MPC and RL code in fixed configurations. The structured format, called \emph{entity-based} in the implementation, retains the identifiers of buildings, equipment, EVs, chargers, flexible loads and requests, together with relations such as building--equipment and charger--EV. It also provides action ranges and availability masks. Dynamic member and asset changes require this structured interface. An installed charger appears as an identified row with its own action.

At each step, the simulator stores the requested action $a_t^{\mathrm{req}}$ and the action $a_t^{\mathrm{app}}$ applied after availability, service windows, electrical limits and command failures are considered. Availability masks and feasible ranges explain which actions were possible; the difference between requested and applied values records what the controller attempted and what the equipment executed.

\subsection{Performance indicators and result exports}
\label{subsec:reward_vs_kpi}
\label{subsec:kpis}

Requirement S2 is that performance indicators and result exports reveal service outcomes as well as aggregate performance. A controller can reduce cost while missing an EV departure, delaying a flexible load, failing a demand-response request or relying on action reduction by the simulator. Reward guides the controller, key performance indicators (KPIs) summarize the run, and exported trajectories show why those indicators changed.

\citylearnvthree retains the energy, cost, emissions, ramping and grid-impact indicators of \citylearnvtwo and adds the families required by D1--D8. \Cref{tab:kpi_families} uses the same names as \cref{tab:requirements}; Appendix~\ref{app:kpis} provides the expanded catalogue.

\begin{table}[!htbp] \centering \caption{Performance-indicator families in CityLearn v3.}
\label{tab:kpi_families}
\scriptsize
\setlength{\tabcolsep}{3pt}
\begin{tabularx}{\linewidth}{L{0.18\linewidth}L{0.41\linewidth}Y}
\toprule
Family & Example metrics & Why it matters \\
\midrule
Energy and power limits & grid import/export, net exchange, peak, ramping, load factor, building/phase-limit exceedance & Retains district-control indicators and shows requests outside configured power limits. \\
Cost and settlement & energy cost, local settlement, comparison cost, savings, demand-response payments and penalties & Separates energy cost, local energy sharing and demand-response payments. \\
Emissions & emissions totals, daily averages and ratios to the reference controller & Keeps environmental performance visible alongside cost and service. \\
PV and local energy sharing & PV generation/export, self-consumption, local import/export, shared energy and sharing ratios & Shows local energy sharing instead of hiding it in aggregate net load. \\
EV service & departure counts, target/minimum/tolerance success, SOC deficit/surplus, charged and V2G energy & Distinguishes grid flexibility from mobility-service failure. \\
Flexible-load service & completed/missed cycles, service level, served/unserved energy and start delay & Evaluates flexible demand as a service with a deadline. \\
Demand-response requests & requests, active timesteps, requested/credited/shortfall energy, compliance, payments and invalid baselines & Measures requested flexibility and its settlement. \\
Data and equipment failures & failures, changed measurements/forecasts/actions, missing values, lost commands and unavailable equipment & Records the conditions under which controller results were produced. \\
Changing community members and assets & active periods, time-stamped changes and changing controller-input/action definitions & Links member and asset changes to valid actions and result periods. \\
Controller inputs and actions & table rows, relations, masks, feasible capacity and requested/applied-action differences & Shows what the controller knew, requested and physically caused. \\
Multiple communities & local rows, summed quantities and weighted ratios & Keeps each community visible next to combined results. \\
Inherited thermal/\allowbreak HVAC/\allowbreak comfort & discomfort ratios, hot/cold deltas, thermal-resilience and unserved-energy indicators & Retains the thermal, HVAC and comfort indicators of CityLearn v2. \\
\bottomrule
\end{tabularx}
\end{table}

\begin{figure}[H]
    \centering
    \includegraphics[width=\linewidth]{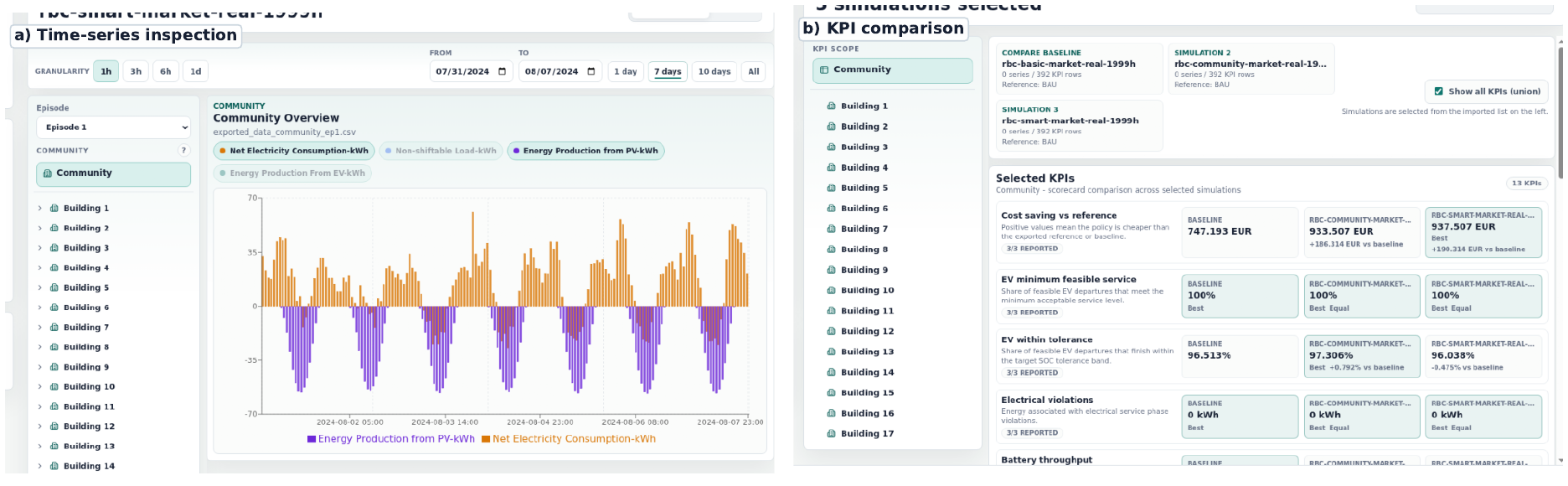}
    \caption{Time-series and KPI comparison views in the CityLearn UI.}
    \label{fig:ui_export_inspection}
\end{figure}

Exports include KPI and time-series files, service summaries, requested/applied-action differences, entity tables, masks, relations and figure inputs. The companion CityLearn UI reads these folders directly; its trajectory and KPI views in \cref{fig:ui_export_inspection} support inspection of the records behind a scorecard.

The scorecard should match the study: EV service and action reductions for mobility, or traded energy, settlement and participant costs for local sharing. Compact scorecards support comparison, while the exports retain the full indicator families for further analysis.

\subsection{Reference controllers and fair comparisons}
\label{subsec:bau}

Requirement S3 is a declared reference case evaluated under the same scenario and accounting rules. Matching service obligations, settlement and electrical limits provides a consistent basis for normalized comparisons between policies.

The reference controller must therefore state how it charges EVs, starts mandatory flexible loads, operates batteries and reacts to demand-response requests or failures. It uses the same settlement and building/phase-limit rules as the evaluated controller.

\begin{table}[!htbp]
\centering
\caption{Reference controllers used in the CityLearn v3 evaluation workflow.}
\label{tab:baseline_categories}
\scriptsize
\setlength{\tabcolsep}{3pt}
\begin{tabularx}{\linewidth}{L{0.20\linewidth}L{0.25\linewidth}Y}
\toprule
Policy class & What it represents & Interpretation in REC studies \\
\midrule
\texttt{NormalNoBatteryPolicy} & Day-to-day operation for EVs and flexible loads without stationary battery control. & Separates native service demand from the value of stationary storage. \\
\texttt{NormalPolicy} & BAU operation with the same native service logic plus simple PV self-consumption storage. & Serves as the main reference line for annual scorecards and normalized KPI interpretation. \\
\texttt{RBCBasicPolicy} & Service-aware rule-based control with simple price response. & Checks whether a transparent controller can improve cost or import while tracking EV targets more precisely than normal charging. \\
\texttt{RBCSmartPolicy} & Solar-, price- and peak-aware rule-based control. & Provides a stronger internal RBC reference for grid, carbon and service KPIs. \\
\texttt{RBCCommunityPolicy} & REC-aware rule-based control using community surplus and import context. & Tests whether local sharing and settlement objectives can be inspected separately from grid and carbon outcomes. \\
\bottomrule
\end{tabularx}
\end{table}
\FloatBarrier

KPIs can then be reported in raw form and, when appropriate, relative to the declared reference controller $b$. A generic ratio and difference are
\begin{align}
    \kappa^{\mathrm{ratio}}_j(\pi,b) &= \frac{\kappa_j(\pi)}{\kappa_j(b)}, \\
    \Delta\kappa_j(\pi,b) &= \kappa_j(\pi)-\kappa_j(b),
    \label{eq:kpi_normalization}
\end{align}
where $\pi$ is the evaluated controller, $b$ is the reference controller and $\kappa_j$ is indicator $j$. The ratio is defined only when $\kappa_j(b)\neq 0$. Service failures, invalid demand-response baselines, shortfalls and power-limit exceedances remain visible as raw counts, rates or gates instead of being hidden by normalization.

\subsection{Software validation}
\label{subsec:validation_traceability}

Requirement S4 concerns checks on scenario rules, interfaces and reported results. \citylearnvthree combines regression tests with targeted REC tests, time-unit and physical audits, controller-input checks, BAU/KPI consistency checks and thermal/HVAC compatibility tests. \Cref{tab:validation_evidence} reports their scope and observed outcomes. The checks were run in June 2026 using Python 3.10.12. The test logs and machine-readable audit outputs are retained with the manuscript.

\begin{table}[!htbp] \centering \caption{Software validation checks used in this manuscript revision.}
\label{tab:validation_evidence}
\scriptsize
\setlength{\tabcolsep}{2.5pt}
\begin{tabularx}{\linewidth}{L{0.24\linewidth}L{0.47\linewidth}Y}
\toprule
Check & Protected failure mode & Current evidence \\
\midrule
Full regression suite & General simulator behaviour and compatibility with existing CityLearn tests. & 423 passed tests; 18 warnings. \\
Targeted REC requirements & KPI exports, demand-response requests, data and equipment failures, structured controller inputs, changing community members and assets, and runs with multiple communities. & 103 passed tests; 8 warnings. \\
Physics and time-unit audit & Energy/power consistency when the same scenario is evaluated at 60, 300, 900 and 3600 s timesteps. & 16/16 scenarios passed. \\
Controller-input and change audit & Persistent member/equipment identities, relations, action rows and configuration versions when equipment or members change. & Strict pass. \\
BAU/KPI sanity audit & Reference-controller rows, duplicated KPI keys, finite KPI values and delta-to-BAU identities in controller exports. & 3317 rows per export, 1401 BAU-derived rows and no duplicate KPI keys. \\
Thermal/HVAC compatibility tests & Inherited thermal and heat-pump behaviour alongside the new REC capabilities. & 12 passed tests. \\
\bottomrule
\end{tabularx}
\end{table}

Each figure in \cref{sec:applications} is linked to its scenario descriptor, command, raw output folder, KPI table and figure input. These records identify the timestep, controller inputs, reference controller, events and KPI scope behind each plot.

\FloatBarrier \subsection{Computational cost}
\label{subsec:performance_contract}

Requirement S5 concerns the cost of repeated controller experiments. High-frequency data increase storage and loading requirements, structured inputs enlarge the returned data, and synchronized communities increase the state and indicators processed at each step.

The comparisons in \cref{tab:runtime_storage_evidence} were run on an Intel Core i9-13980HX processor (24 cores, 32 threads), with 29.1 GiB of RAM available to Linux, using Python 3.10.12 and Linux 6.8.0-124 on x86\_64. Control actions were zero and rendering was disabled. Step durations were measured around \texttt{env.step()} using \texttt{perf\_counter}, excluding controller decision time. The basic flat/structured comparison reports means over five process-isolated runs per configuration (seeds 0--4), each with 599 executed hourly steps. The detailed-input and multi-community cases use one 599-step hourly run per configuration; the CSV/Parquet cases use one 239-step run per configuration at 15 s resolution. Initialization and reset were timed separately.

\par{}

\begin{table}[!htbp] \centering \caption{Computational-cost evidence.}
\label{tab:runtime_storage_evidence}
\scriptsize
\setlength{\tabcolsep}{2.5pt}
\begin{tabularx}{\linewidth}{L{0.19\linewidth}L{0.25\linewidth}L{0.25\linewidth}Y}
\toprule
Check & Compared cases & Observed change & Interpretation \\
\midrule
Basic structured input & Matched building/phase-limit run, flat input versus the basic structured input. & 5.30 to 5.51 ms/step; about +0.21 ms/step or +4.1\%. & Persistent identities, relations, masks and action feedback are available with measured but small basic overhead. \\
High-frequency storage & Two 15 s CSV/Parquet pairs: power limits and changing community members and assets. & 2.446 GB to 12.405 MB and 2.619 GB to 12.740 MB; initialization falls from 25.37 to 15.61 s and from 28.59 to 18.92 s. & Parquet makes both datasets about 200 times smaller and reduces initialization time by 34--38\%; post-load step time remains similar. \\
Structured-input size & Flat input versus the most detailed standard structured input. & 4.88 to 10.18 ms/step; 165 flat values to 2538 table values and 302 feature names. & Step time approximately doubles while the exposed input payload becomes more than 15 times larger. \\
Multiple communities & One, two and four synchronized communities in the heterogeneous run. & 5.02, 7.08 and 11.41 ms/step; four communities export 5227 KPI rows, including 107 combined rows. & Four synchronized communities require 2.27 times the single-community step time while retaining local and combined outcomes. \\
\bottomrule
\end{tabularx}
\end{table}
\FloatBarrier

The basic structured interface adds 4.1\% to mean step time. The most detailed input exposes over 15 times as many values at 2.09 times the step time. Four communities average 11.41 ms/step, approximately 88 steps/s, with 5227 local and combined KPI rows.

\section{Application examples}
\label{sec:applications}

The application examples examine how scenario conditions affect service delivery, visible peaks, settlement and controller records. \Cref{tab:application_protocol} summarizes the questions and experimental configurations.

The scenario configurations are summarized in Appendix~\ref{app:datasets}.

\begin{table}[H]
\centering
\caption{Application-example protocol used in this manuscript.}
\label{tab:application_protocol}
\scriptsize
\setlength{\tabcolsep}{3pt}
\begin{tabularx}{\linewidth}{L{0.20\linewidth}L{0.24\linewidth}L{0.25\linewidth}Y}
\toprule
Example & Study question & Scenario/controller & Reported outcome \\
\midrule
Reference controllers & Can BAU and transparent reference controllers be interpreted with the expanded KPI families on a full annual REC/EV scenario? & 8760-step hourly REC scenario with EVs, flexible loads, local energy sharing and building/phase power limits; legacy EV-RBC, BAU and rule-based controllers. & The scorecard compares energy, cost, EV and flexible-load service, settlement and requested power-limit exceedances. \\
Time resolution and energy units & Can the same energy trace imply different visible power peaks at different time resolutions? & One synthetic annual 15 s REC trace viewed at 1 h, 15 min and 15 s; all views are derived from the same physical signal. & Equal annual energy can produce different visible peaks, so the declared timestep matters for capacity-sensitive studies. \\
Local energy sharing and settlement & Does local settlement change the annual economic interpretation of a community-aware controller? & Full-year \texttt{RBCCommunityPolicy}; grid-only comparison versus configured local settlement for the same trajectory. & Local exchange changes final cost and member outcomes and cannot be inferred from district net load alone. \\
Demand-response requests & Can a time-window request be translated into delivery and settlement evidence? & 180-step scenario with three non-overlapping requests; one evening request shown against operation without a request. & Request-level results expose requested energy, credited response, shortfall and payment. \\
Changing community members and assets & Do time-stamped changes propagate to controller inputs, actions and results? & One run in which a member, chargers and PV equipment are added or removed; detailed structured-input results are archived. & Member/equipment counts and controller rows change together while identities remain traceable. \\
Data and equipment failures & Do repeatable failures change the controller trajectory and appear in the results? & Same controller and period in a clean run and a run with declared measurement, forecast, command and equipment failures. & Declared failures produce identifiable changes in controller inputs, applied actions and exported results. \\
Multiple communities & Can heterogeneous communities be evaluated together without hiding individual outcomes? & Four heterogeneous communities evaluated over the same time window. & Individual profiles and combined outcomes quantify simultaneous surplus and demand across communities. \\
\bottomrule
\end{tabularx}
\end{table}

\subsection{Reference controllers and scorecard interpretation}
\label{subsec:application_scorecard}

The first example combines energy, cost, service, settlement and electrical-constraint indicators in an annual scorecard. It compares the reference policies defined in \cref{subsec:bau} across grid, economic and service outcomes.

The hourly three-phase REC/EV scenario contains EV chargers, one flexible-load service, local sharing and building/phase limits. Its annual window has 8760 timesteps and 8759 executed control steps. Six policies are compared: the older \texttt{Legacy EV-RBC} and the five reference policies in \cref{tab:baseline_categories}. \texttt{NormalPolicy} supplies the BAU reference, combining ordinary EV and flexible-load service with simple PV self-consumption storage.

\begin{figure}[H]
    \centering
    \includegraphics[width=0.98\linewidth]{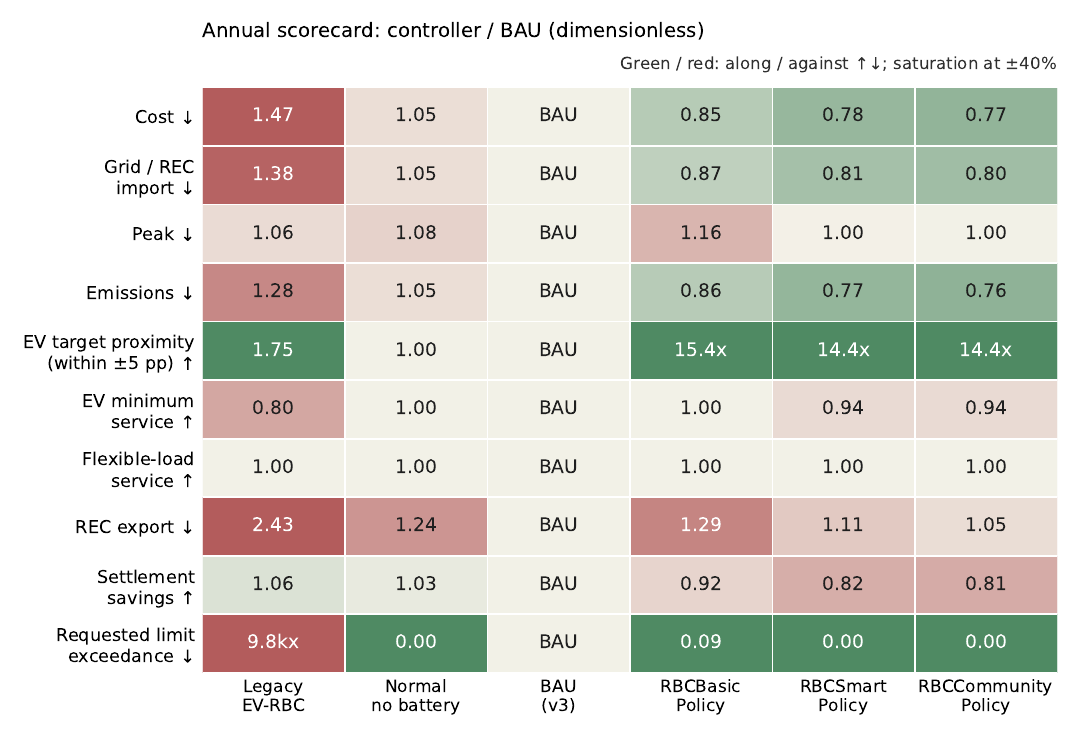}
    \caption{Annual REC/EV controller scorecard normalized to the \citylearnvthree BAU reference.}
    \label{fig:controller_scorecard}
\end{figure}

\Cref{fig:controller_scorecard} gives each indicator as a ratio to BAU, with arrows indicating the comparison direction. EV target proximity measures the proportion of eligible departures within $\pm5$ percentage points of the requested SOC, while minimum service measures attainment of the configured minimum acceptable SOC. The six policies share 2757 eligible departures, the same settlement rules and the same building/phase limits. BAU attains the minimum in all eligible departures but lies within the symmetric target band in 6.3\%, because its charging rule aims for 100\% SOC and frequently exceeds the requested target.

Legacy EV-RBC produces an annual cost of 42.2 kEUR, grid import of 230.1 MWh and emissions of 38.4 tCO$_2$, compared with 28.7 kEUR, 166.8 MWh and 30.0 tCO$_2$ for BAU. It meets the minimum feasible EV target for 80.1\% of feasible departures and accumulates 49.5 MWh of requested power-limit exceedance energy.  

\texttt{RBCBasicPolicy} reduces annual cost to 24.4 kEUR and grid import to 145.4 MWh while retaining 100\% minimum EV service and reaching 98.0\% target proximity. Its peak rises from 104.7 kW for BAU to 121.4 kW. \texttt{RBCSmartPolicy} reduces cost to 22.3 kEUR, import to 135.2 MWh and emissions to 23.1 tCO$_2$. \texttt{RBCCommunityPolicy} reaches 22.0 kEUR, 133.3 MWh, a 104.6 kW peak and 22.8 tCO$_2$. The two smart policies attain target proximity of 91.4\% and 91.2\%, respectively, with minimum EV service of 93.8\% for both and flexible-load service of 100\%. Among the RBC policies, Basic delivers full minimum EV service, while Smart and Community achieve the lowest cost and import. The target-proximity ratios compare the frequency of departures within the requested SOC band. Settlement savings reflect the timing and volume of local exchange.

\begin{figure}[H]
    \centering
    \includegraphics[width=0.98\linewidth]{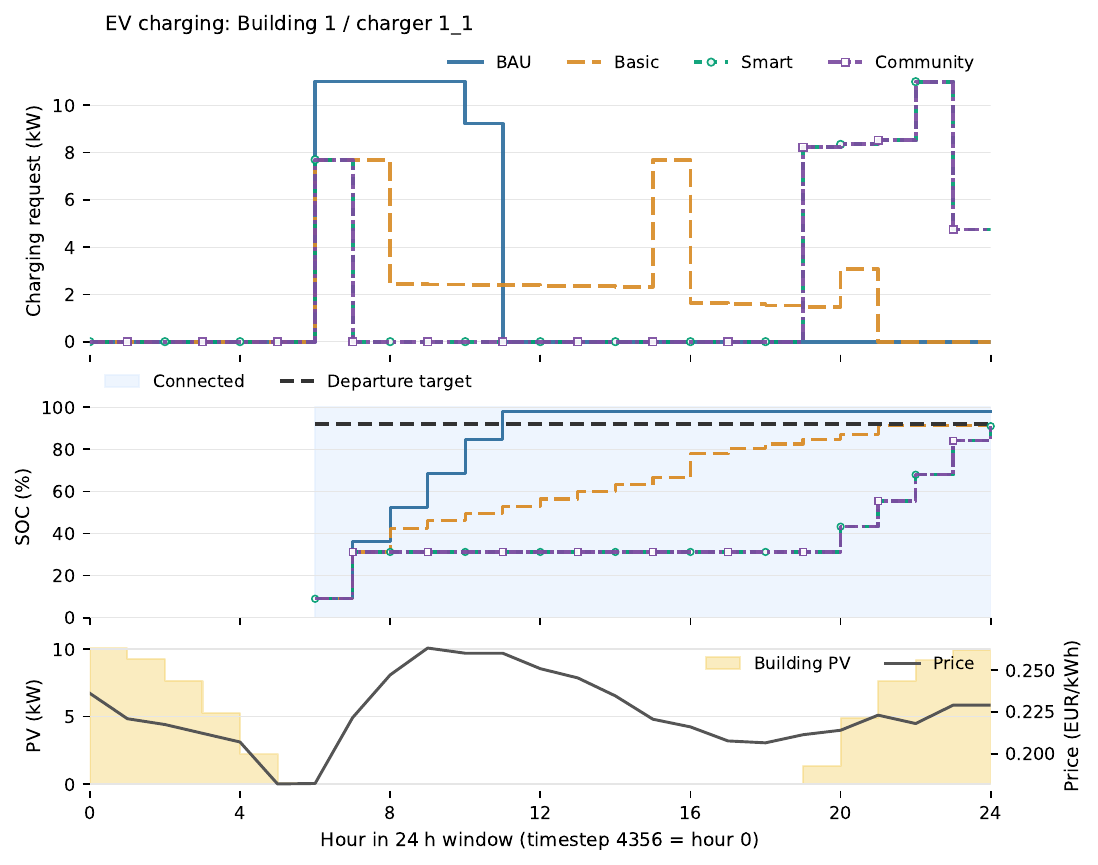}
    \caption{EV charging requests over 24 h and the corresponding timestep-boundary SOC states for one charger.}
    \label{fig:baseline_daily_ev_behavior}
\end{figure}

\Cref{fig:baseline_daily_ev_behavior} follows one charger through a 24 h window. BAU begins charging when the vehicle connects, while the RBC policies shift part of the request to later hours with different PV output and prices. The middle panel shows connection status, requested departure SOC and the SOC at timestep boundaries, including the state reached after the final control action. Smart and Community finish at 91.0\% SOC against a 92\% target, within the configured tolerance. The lower panel supplies the building's PV and price context.

\subsection{Time resolution and hidden peak visibility}
\label{subsec:example_frequency}

Short timesteps retain brief EV-charging and process-load peaks that hourly averaging can conceal. This experiment uses a deterministic replay of one synthetic annual 15 s REC trace and constructs 15 min and 1 h views by averaging the same signal. All views retain 349.0 MWh of load, 147.6 MWh of PV generation and 19.3 MWh of EV charging, isolating the effect of temporal aggregation on visible power.

\begin{figure}[H]
    \centering
    \includegraphics[width=0.92\linewidth]{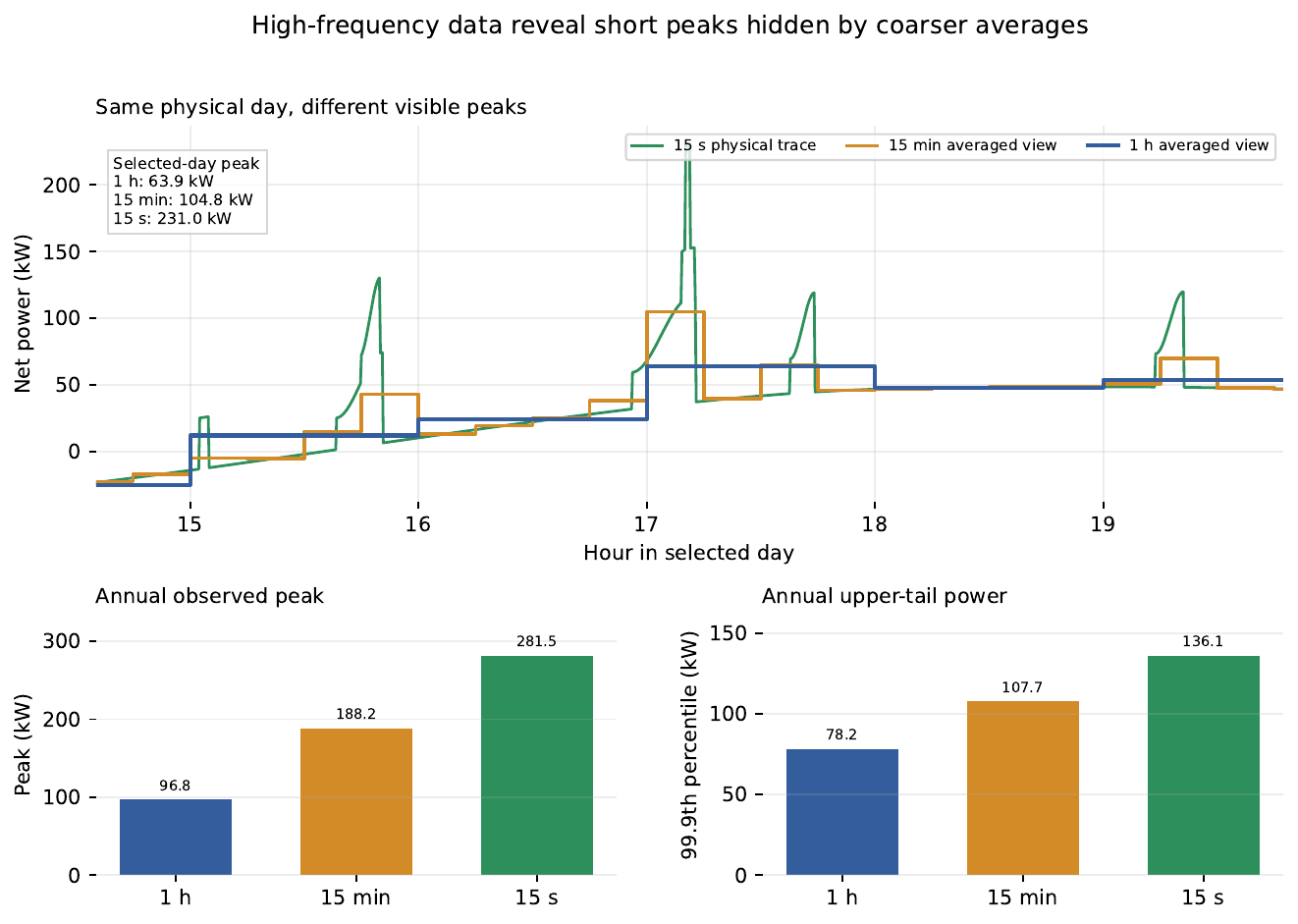}
    \caption{High-frequency peak visibility from one synthetic 15 s trace.}
    \label{fig:high_frequency_bau}
\end{figure}

\Cref{fig:high_frequency_bau} shows the consequence on a selected day and over the annual trace. On the selected day, the native 15 s trace reaches 231.0 kW when a short high-power event occurs. The same day reaches only 104.8 kW in the 15 min averaged view and 63.9 kW in the hourly view, so the hourly representation hides most of the short event. Over the annual trace, the hourly view reports an observed peak of 96.8 kW and a 99.9th percentile power of 78.2 kW. The 15 min view reports 188.2 kW and 107.7 kW, while the native 15 s view reports 281.5 kW and 136.1 kW. Thus, the same annual energy balance can imply substantially different visible peaks and upper-tail behaviour depending on the timestep used by the controller and evaluator.

Aggregation conceals short capacity-relevant events even when annual energy is unchanged. The declared timestep therefore affects the peaks and upper-tail power available to a controller or evaluator.

\subsection{Local energy sharing and settlement}
\label{subsec:example_rec_electrical}

This experiment values the annual \texttt{RBCCommunityPolicy} trajectory from \cref{subsec:application_scorecard} under grid-only accounting and configured REC settlement. Both valuations use the same import, export, PV, battery and service trajectories over 8760 hourly timesteps, with 8759 executed control steps.

 Under grid-only accounting, the annual electricity cost of the \texttt{RBCCommunityPolicy} trajectory is 24.8 kEUR. With local settlement, the same trajectory is valued at 22.0 kEUR, a reduction of 2.8 kEUR or 11.3\%. The run records 17.0 MWh of locally traded energy, 133.3 MWh of residual grid import and 48.3 MWh of residual grid export. Local exchange covers 11.3\% of demand and 26.1\% of export energy under the configured rule. All 17 members reduce annual cost, with a median saving of 121 EUR and a range from 68.8 to 525.7 EUR. These savings accompany 91.2\% EV target proximity, 93.8\% minimum EV service and 100\% flexible-load service for the same controller.

\begin{figure}[H]
    \centering
    \includegraphics[width=0.96\linewidth]{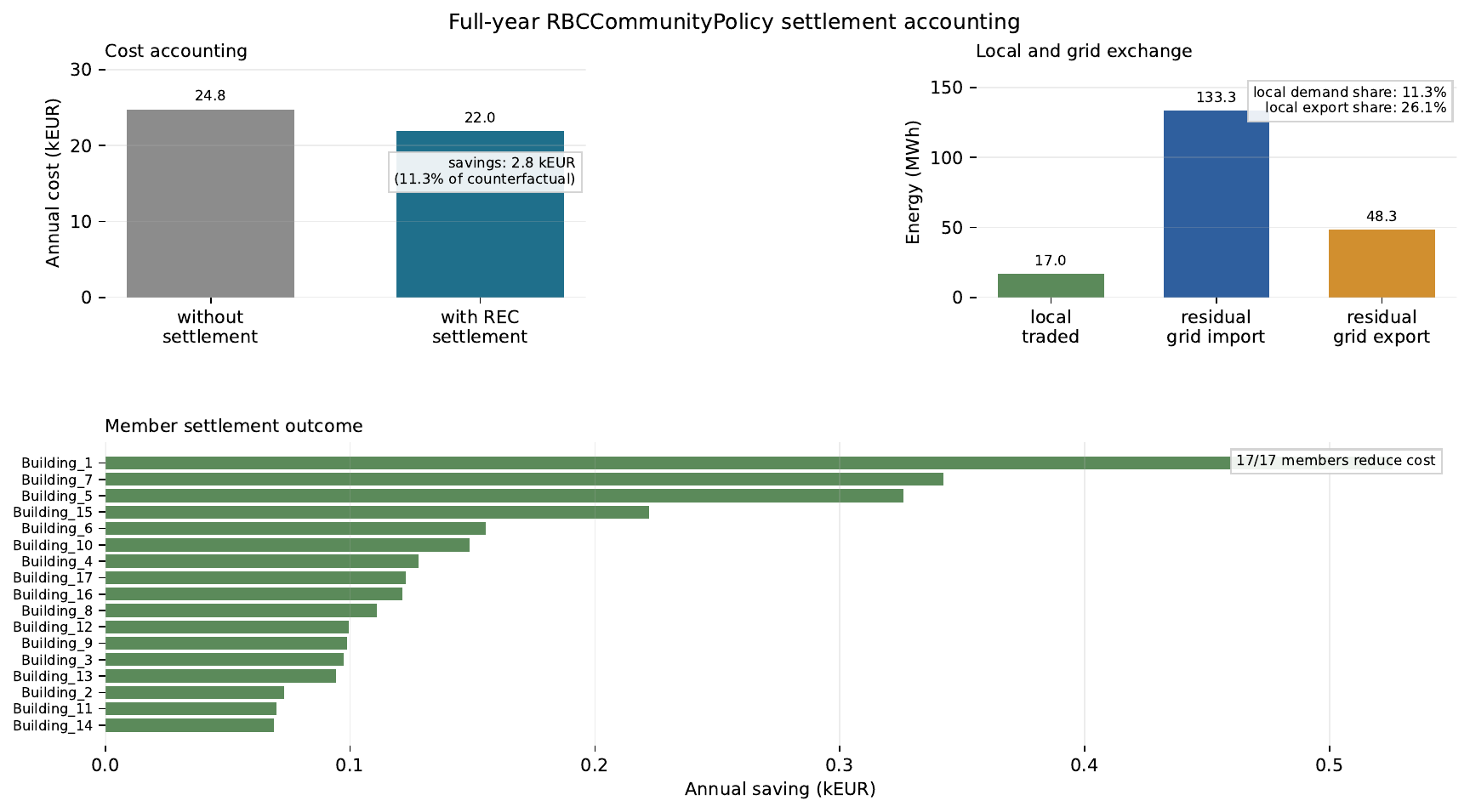}
    \caption{Full-year REC settlement accounting for the same \texttt{RBCCommunityPolicy} trajectory.}
    \label{fig:rec_settlement_full_year}
\end{figure}

The range of member savings shows how local settlement distributes community-level benefits. The participant-level exports provide the quantities needed to compare allocation rules and fairness, including the authors' related simulator-based analysis \citep{fonseca2026_rec_equity}.

\subsection{Demand-response delivery and settlement}
\label{subsec:example_dr}

This experiment applies an event-aware dispatch to explicit demand-response requests and tracks delivery, credit, shortfall and settlement.

\begin{figure}[H]
    \centering
    \includegraphics[width=0.92\linewidth]{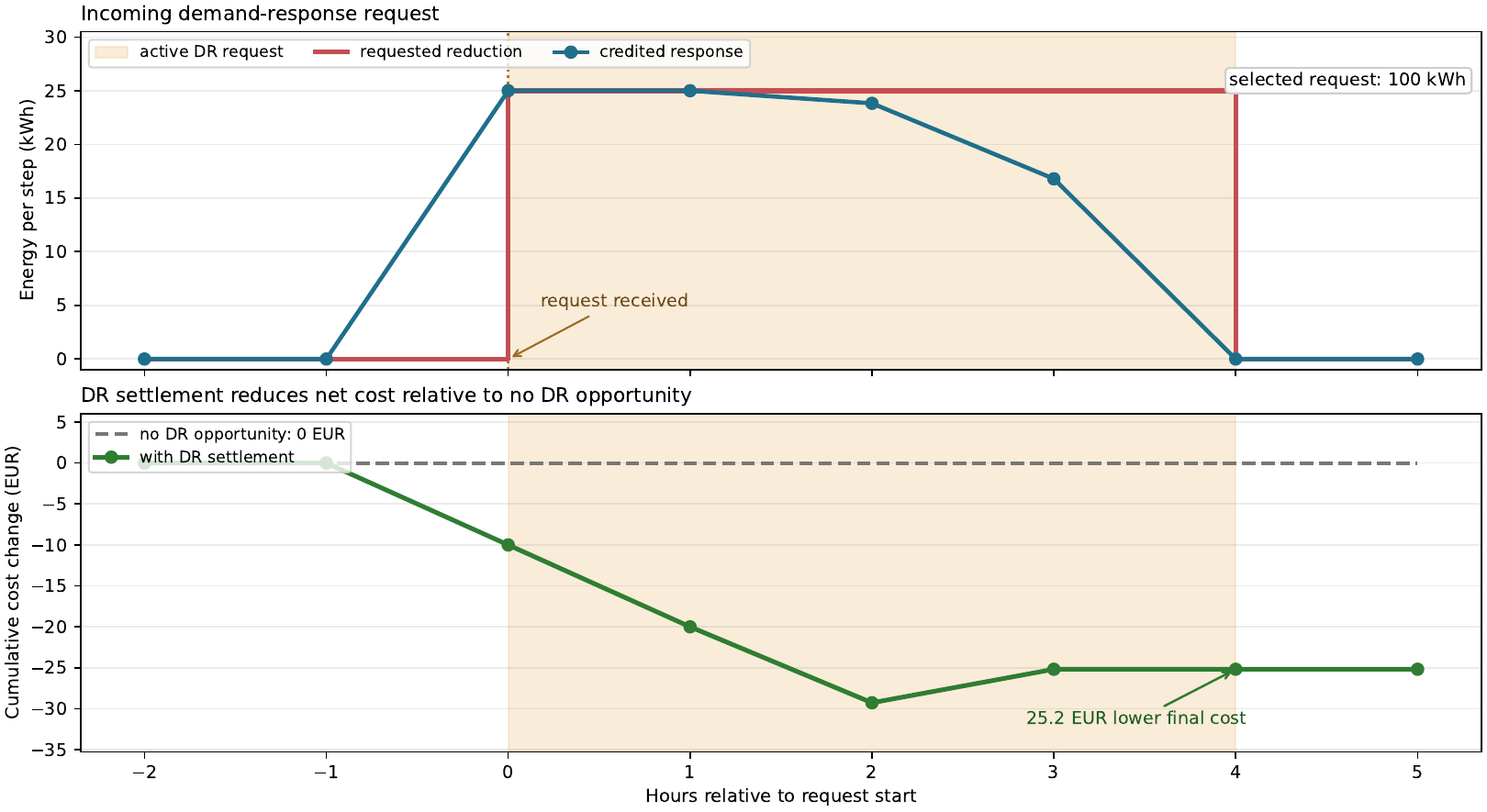}
    \caption{Selected demand-response request and settlement opportunity.}
    \label{fig:dr_delivery}
\end{figure}

The configured 180-step scenario contains three non-overlapping requests, active over 10 hourly timesteps and requesting 190.0 kWh in total. \Cref{fig:dr_delivery} focuses on one evening down request. In the upper panel, the shaded region marks the active window. The red step curve asks for 25 kWh of reduction per timestep, or 100.0 kWh in total. The blue markers show the response credited for payment. It follows the request closely for the first three timesteps and falls below it in the final timestep, creating a shortfall.

The lower panel compares settlement with a no-request counterfactual, whose DR-related cost change remains zero. The green curve falls as credited response is remunerated; negative values indicate lower net cost relative to that counterfactual. A shortfall penalty reduces the benefit near the end of the event. The selected request finishes with 90.6 kWh of credited response, 7.4 kWh of shortfall and 25.2 EUR lower net cost.

Across the three requests in \cref{tab:dr_three_request_aggregate}, event-aware dispatch reduces total shortfall from 101.1 to 51.6 kWh, raises compliance from 0.43 to 0.70 and lowers cost after settlement by 73.0 EUR relative to reference operation. Net revenue remains negative after shortfall penalties. Delivered energy records all movement in the requested direction, while compliance uses credit capped at each timestep. Credit and penalized shortfall use different thresholds because tolerance reduces the latter. These records connect the change in net cost to the response delivered during each request.

\begin{table}[!htbp]
\centering
\caption{Combined demand-response results for the three-request scenario.}
\label{tab:dr_three_request_aggregate}
\scriptsize
\setlength{\tabcolsep}{3pt}
\begin{tabular}{lrrrrrr}
\toprule
Case & Requested & Delivered & Shortfall & Compliance & Net DR revenue & Cost after settlement \\
 & kWh & kWh & kWh & ratio & EUR & EUR \\
\midrule
Reference operation & 190.0 & 82.5 & 101.1 & 0.43 & -100.2 & 863.6 \\
Event-aware dispatch & 190.0 & 300.9 & 51.6 & 0.70 & -16.0 & 790.6 \\
\bottomrule
\end{tabular}
\end{table}
\FloatBarrier

\subsection{Changing community members and assets}
\label{subsec:example_topology}

This example tests whether member and equipment changes propagate to the structured controller input within one continuous run.

\begin{figure}[H]
    \centering
    \includegraphics[width=0.86\linewidth]{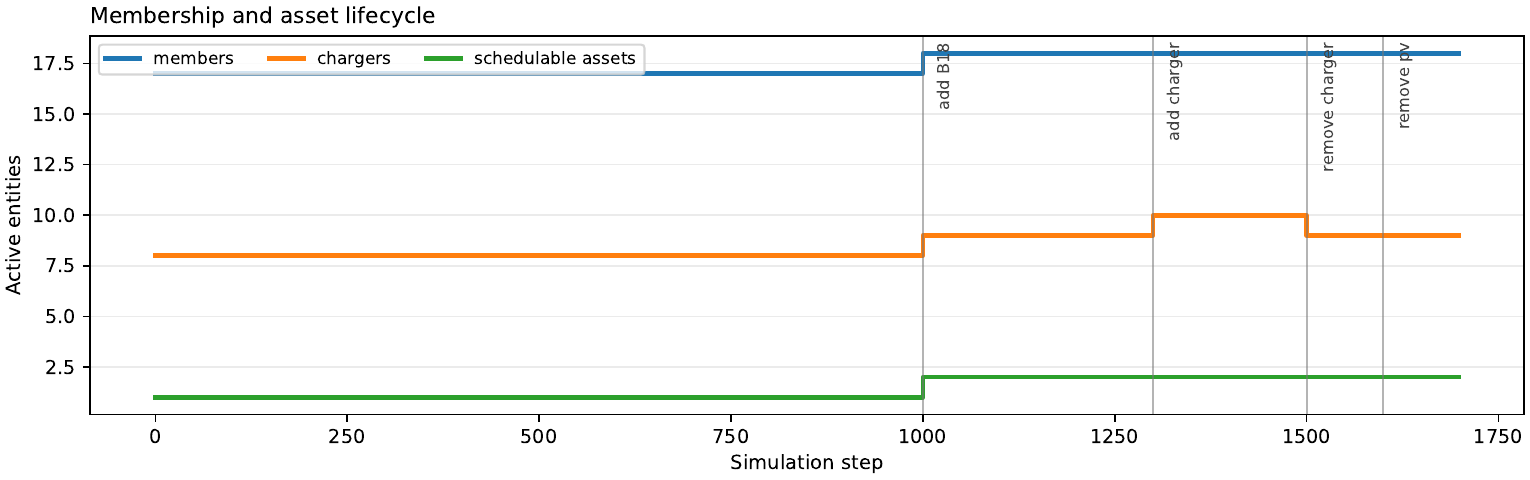}
    \caption{Counts of changing community members and assets.}
    \label{fig:dynamic_membership_lifecycle}
\end{figure}

Four changes occur in one run: Building 18 is added at timestep 1000, a charger is added to Building 2 at 1300, a charger is removed from Building 5 at 1500 and PV is removed from Building 11 at 1600. \Cref{fig:dynamic_membership_lifecycle} shows the resulting counts. Active members increase from 17 to 18, active chargers increase from 8 to 10 and then return to 9, and active flexible loads increase from 1 to 2 when the new member enters.

The structured input changes at the same timesteps: identified rows increase from 67 to 74 and end at 71, relations move from 74 to 84 and then 80, and action rows move from 26 to 30 and then 29. The largest input contains 2295 table values. Persistent identifiers allow these changes to be traced to individual members and assets. Archived input sets range from 85 feature names in the basic set to 302 in the largest standard set.

\subsection{Data and equipment failures}
\label{subsec:example_robustness}

\par{}

The same hourly week is run twice with a storage policy: once with unchanged inputs and equipment availability, and once with declared failures. The policy produces non-zero storage requests around the failure windows so that lost commands and unavailable equipment leave visible traces in the trajectory and service records.

\begin{figure}[H]
    \centering
    \includegraphics[width=0.94\linewidth]{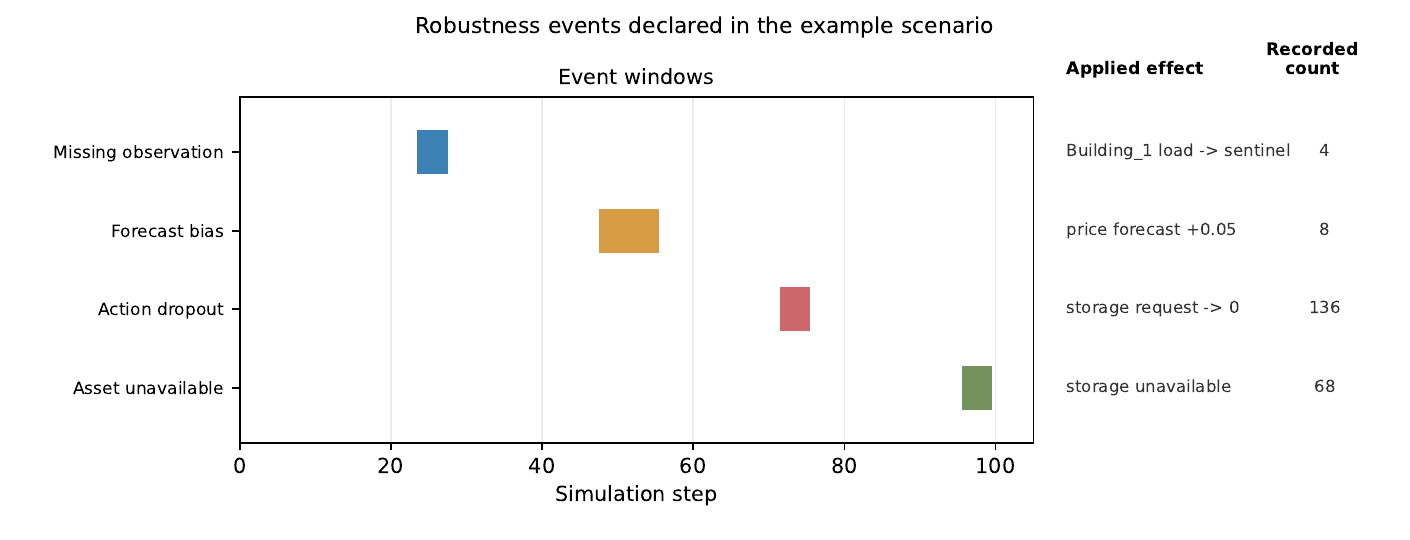}
    \caption{Data and equipment failures in the example scenario.}
    \label{fig:robustness_events}
\end{figure}

The file declares four failures over 20 active timesteps: missing building-load measurements, biased electricity-price forecasts, lost storage commands and temporary storage unavailability. \Cref{fig:robustness_events} places them on the same simulation axis and reports the associated counts. The missing-meter interval creates 4 records, the forecast-bias interval creates 8, the command failure affects 136 action entries across the storage actions, and the equipment failure records 68 unavailable equipment--timestep pairs. The figure shows what was declared, when it occurred and how it was counted.

\begin{figure}[H]
    \centering
    \includegraphics[width=0.94\linewidth]{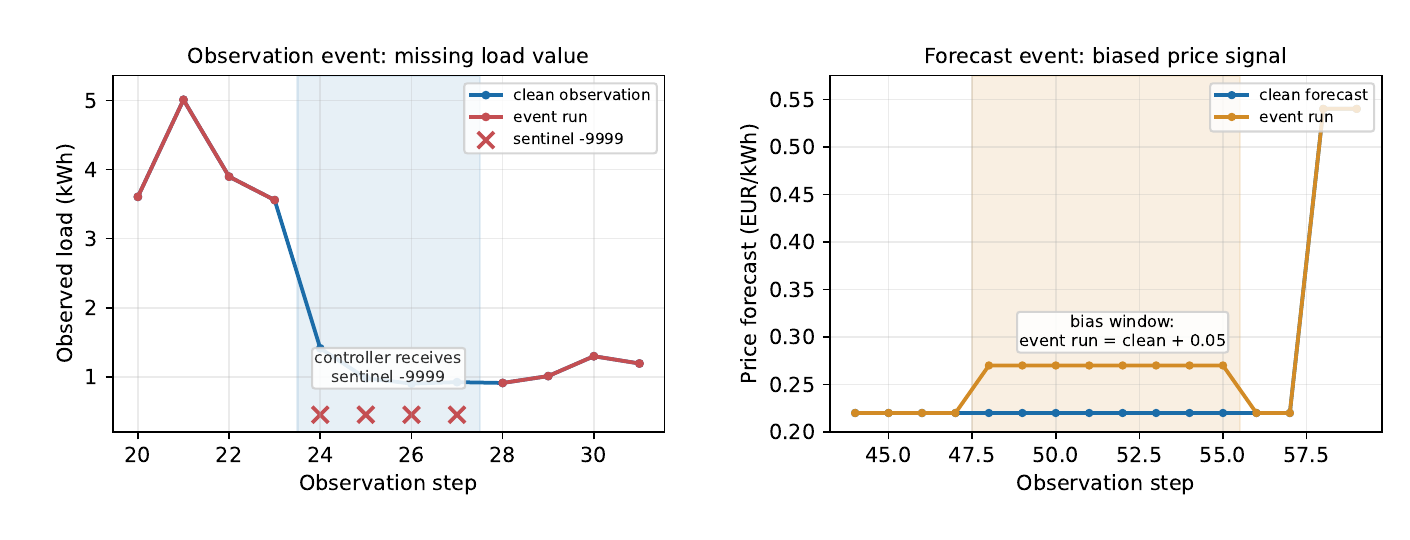}
    \caption{Missing measurements and biased forecasts.}
    \label{fig:robustness_data_events}
\end{figure}

\Cref{fig:robustness_data_events} shows the effects of measurement and forecast failures. On the left, the clean trace shows the physical meter value. During the failure, the controller-visible value becomes the configured missing-value marker, \(-9999\), rather than a physical zero. The crosses at the bottom mark the missing-measurement interval. On the right, affected forecast values are shifted by \(+0.05\) EUR/kWh. The controller thus receives corrupted information while the export records when and how it was changed.

The same run also records lost commands and unavailable equipment. During the command-failure window, requested storage actions remain recorded while the corresponding applied actions become zero; mean net electricity is then 25.33 kWh/step higher than in the clean run. Over the full run, the grid-import and cost ratios increase by approximately 0.011 and 0.0041, while the peak ratio is unchanged. The event records link these changes to the affected measurement, forecast, command and equipment channels.

\subsection{Multiple communities}
\label{subsec:example_multi}

Synchronized experiments can reveal complementary demand and generation profiles while retaining each community's objectives and results.

The experiment evaluates four REC scenarios over the same 600-step hourly window. REC 1 has many EVs; RECs 2--4 have different building demand and PV profiles. Each electrically independent community is controlled with \texttt{RBCSmartPolicy}. After the run, post-processing matches simultaneous export in one community with import in another under a common accounting rule. The matched quantities measure inter-community sharing opportunities from the recorded trajectories.

\begin{figure}[H]
    \centering
    \includegraphics[width=0.98\linewidth]{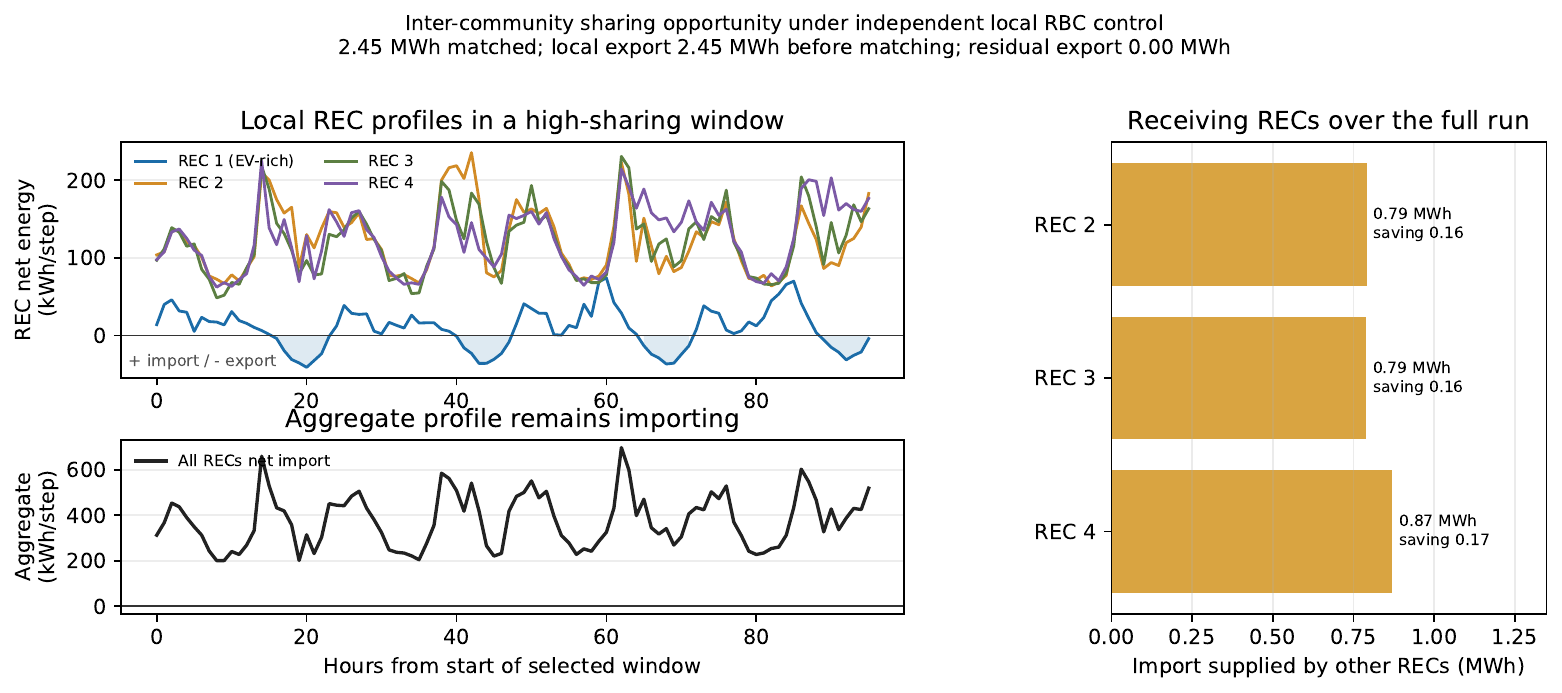}
    \caption{Inter-community sharing opportunity under independent local control.}
    \label{fig:multi_community_exchange}
\end{figure}

\Cref{fig:multi_community_exchange} shows the result in two complementary ways. The left side plots a high-sharing window. REC 1 repeatedly exports while the other RECs remain importers, yet the aggregate profile of all four RECs is still positive. The separate REC profiles reveal simultaneous local surplus and demand beneath a positive aggregate net load.

The right side gives the full-run accounting view. Over the 600-step window, 2.45 MWh of local export is matched with demand in other communities, and no residual export remains after this matching calculation. Of that matched demand, 0.79 MWh of REC 2 import, 0.76 MWh of REC 3 import and 0.87 MWh of REC 4 import are matched mainly with REC 1 export; small additional matches go from the other RECs to REC 3.

Under the normalized accounting rule, locally matched energy is valued at 80\% of the retail import price and grid export remuneration is zero. This gives recipient-side savings of 0.16, 0.16 and 0.17 normalized cost units for the three receiving communities, and 0.49 normalized cost units in total. The combined records quantify the potential savings associated with simultaneous surplus and demand across the four communities.

\section{Discussion and conclusions}
\label{sec:discussion}
\label{sec:conclusion}

This paper presented \citylearnvthree as a configurable framework for studying control in renewable energy communities whose participants, equipment and operating conditions can change over time. Scenario configuration determines both what the controller observes and how its decisions are applied, with service requirements and electrical limits carried through to evaluation. Building and phase active-power limits are enforced alongside timestep energy balances, while local sharing and settlement connect physical operation to participant costs. This extends the building-energy and thermal modelling capabilities of \citylearnvtwo into a common environment for realistic REC control experiments.

The application results demonstrate why energy, service and economic indicators need to be read together. In the annual controller comparison, lower electricity costs and grid imports coincide with different levels of EV departure service. Local settlement, in turn, changes participant costs without changing the physical trajectory, and demand-response remuneration depends on both delivered response and shortfall. Temporal resolution also shapes the interpretation of a trajectory: averaging the same 15 s trace to 15 min or 1 h preserves energy totals but conceals short power peaks.

For controller research, the value of this framework lies in being able to vary those conditions while retaining a consistent link between decisions and outcomes. Requested and applied actions explain how electrical limits or equipment availability affect operation, and service records identify the consequences for community members. The open-source implementation and public example scenarios provide a basis for extending these studies to new policies and configurations. CityLearn v3 thus supports comparisons that connect energy and economic performance to the decisions, constraints and delivered services that produce it.

\section*{Acknowledgements} This work was supported by Fundação para a Ciência e a Tecnologia, I.P. (FCT), under the supervision and superintendence of the Ministry of Education, Science and Innovation, through doctoral grant 2024.00855.BD, financed by the Portuguese State Budget and co-financed by the European Social Fund Plus (FSE+) through the Programa Demografia, Qualificações e Inclusão (PDQI/PESSOAS 2030). This work was also supported by the DEMFLEX operation (COMPETE2030-FEDER-01657200 -- 19215), supported by the Innovation and Digital Transition Programme (COMPETE 2030), under Portugal 2030, and co-financed by the European Union. The views and opinions expressed in this paper are those of the author alone and do not necessarily reflect the views of FCT, the European Union, KDT JU, or the respective funding and granting authorities. Neither these institutions nor the granting authorities can be held responsible for them.

\section*{Disclosure statement} No potential conflict of interest was reported by the authors.


\section*{Data and code availability statement}
CityLearn v3 source code and public example scenarios are available from the official repository, \url{https://github.com/citylearn-project/CityLearn}.  

\bibliographystyle{tfcad}
\bibliography{references}

\appendix

Appendix~\ref{app:datasets} records the validation and application configurations; Appendix~\ref{app:kpis} expands the KPI definitions; and Appendix~\ref{app:implementation_map} maps implementation areas to the capabilities discussed in the paper.

\section{Scenario evidence and configuration reference}
\label{app:datasets}

\subsection{Scenarios used as evidence}

{\scriptsize
\setlength{\tabcolsep}{3pt}
\begin{longtable}{@{}p{0.12\linewidth}p{0.15\linewidth}p{0.21\linewidth}p{0.14\linewidth}p{0.24\linewidth}@{}}
\caption{Software-validation configurations and application-evidence scenarios.}\label{tab:dataset_descriptor}\\
\toprule
Scenario & Temporal scope & Equipment represented & Changing conditions & Paper role \\
\midrule
\endfirsthead
\toprule
Scenario & Temporal scope & Equipment represented & Changing conditions & Paper role \\
\midrule
\endhead
Base REC & \SI{3600}{\second}; 48-step KPI check, 600-step runtime check and full-year reference-controller scorecard & Buildings, PV, batteries, EV chargers and flexible loads & none & BAU/KPI checks and the annual controller scorecard. \\
EV departure target & \SI{3600}{\second}; scorecard service result & EVs, chargers, PV, batteries and buildings & none & Checks strict target, tolerance and minimum-feasible EV departure indicators. \\
Flexible-load schedules and deadlines & \SI{3600}{\second}; scorecard service result & Buildings and flexible appliances & Start windows and deadlines & Checks completed/missed cycles and flexible-load service indicators. \\
Demand-response requests & \SI{3600}{\second}; 48-step KPI check and 180-step application experiment & Buildings, PV, batteries and EV chargers & Demand-response request file & Checks request, delivery, shortfall and settlement indicators. \\
Data and equipment failures & \SI{3600}{\second}; 48-step KPI check and hourly-week application experiment & Buildings, PV, batteries and EV chargers & Data and equipment failure file (\texttt{robustness} internally) & Checks measurement, forecast, command and equipment-availability failures. \\
Changing community members and assets & 60, 300, 900 and \SI{3600}{\second}; 160-step audits, plus an hourly application with changes at steps 1000--1600 & Buildings, chargers, PV, batteries and flexible loads & Member and equipment changes & Checks identities, active periods, controller inputs and result windows. \\
Multiple communities & \SI{3600}{\second}; 48-step KPI check and 600-step combined run & Four heterogeneous REC scenarios & none in current check & Checks individual and combined KPI rows and matched inter-community energy. \\
15 s Parquet & 15 s; file-size, 240-step storage/runtime check and 24 h BAU comparison & Power-limit and changing-equipment demonstrations & optional & Checks storage footprint, loading and time-resolution-aware BAU accounting. \\ Synthetic annual trace & Native 15 s annual replay and 15 min/1 h averages & Aggregate load, PV and EV traces & None & Quantifies the effect of temporal aggregation on visible power. \\
\bottomrule
\end{longtable}
}

\subsection{Exact configuration reference for Section 3}

\Cref{tab:schema_config_map} lists the scenario keys and linked-file fields for requirements D1--D8 in the order used in \cref{tab:requirements}. Internal names and legacy aliases are included where relevant.  

{\scriptsize
\setlength{\tabcolsep}{3pt}
\begin{longtable}{@{}p{0.18\linewidth}p{0.35\linewidth}p{0.39\linewidth}@{}}
\caption{Exact configuration reference for the eight REC requirements in Section~3.}\label{tab:schema_config_map}\\
\toprule
ID and requirement & Scenario declaration & Linked file or record fields and supported values \\
\midrule
\endfirsthead
\toprule
ID and requirement & Scenario declaration & Linked file or record fields and supported values \\
\midrule
\endhead
D1 Multiple communities & \field{MultiCommunityEnv} receives \field{communities}; each entry contains \field{community_id}, \field{schema}, optional \field{env_kwargs} and optional \field{weight}. Wrapper output uses optional \field{render_directory} and \field{render_session_name}. & \field{community_id} is unique; \field{weight} is finite and non-negative, with at least one positive weight. Child scenarios must use matching \field{seconds_per_time_step}, episode length, interface and central-agent settings. \\
D2 Demand-response requests & Top-level \field{demand_response}: \field{enabled}, \field{requests_file}, \field{baseline_method}, \field{baseline_window_seconds} and \field{allow_overlapping_requests}. The current baseline mode is \field{rolling_pre_event_average}; overlapping requests are not currently supported. & Request-file columns: \field{request_id}, \field{issuer}, \field{direction}, \field{start_time_step}, \field{end_time_step}, \field{target_power_kw}, \field{activation_price_eur_per_kwh}, \field{shortfall_penalty_eur_per_kwh} and optional \field{tolerance_power_kw}. Issuer is \field{dso} or \field{tso}; direction is \field{up} or \field{down}. \\
D3 Local energy sharing and settlement & Top-level \field{community_market}: \field{enabled}, \field{local_price_ratio_to_grid_import}, optional \field{import_member_weights}, \field{kpis.community_local_traded_enabled} and \field{kpis.community_self_consumption_enabled}. & The price ratio is clipped to $[0,1]$; \field{intra_community_sell_ratio} is accepted as a legacy alias. Residual grid export is recorded but has zero remuneration in the current implementation; it has no active configurable settlement price. \\
D4 Changing community members and assets & Top-level \field{topology_mode} and \field{topology_events}. Dynamic changes require \field{topology_mode=dynamic} and the entity controller interface. & Event fields: \field{id}, \field{time_step}, \field{operation}, \field{target_member_id}, \field{target_asset_type}, \field{target_asset_id}, optional \field{source_member_id}, \field{source_asset_id} and \field{overrides}. Operations are \field{add_member}, \field{remove_member}, \field{add_asset} and \field{remove_asset}; asset types are \field{charger}, \field{deferrable_appliance}, \field{pv} and \field{electrical_storage}. \\
D5 Flexible-load schedules and deadlines & Under \field{buildings.<id>.deferrable_appliances.<id>}: \field{cycle_profiles_file}, \field{flexibility_schedule_file} and optional \field{attributes.trigger_threshold}. & Cycle-profile columns: \field{profile_id}, \field{duration_steps}, \field{total_energy_kwh}, \field{load_profile}. Schedule columns: \field{cycle_id}, \field{profile_id}, \field{earliest_start_time_step}, \field{latest_start_time_step}, \field{deadline_time_step}, \field{priority}, \field{must_run}. The \field{must_run} field is the mandatory-service flag described in the main text. \\
D6 Data and equipment failures & Top-level data and equipment failure block (\field{robustness} internally): \field{enabled}, \field{events_file}, \field{random_seed}, \field{missing_replacement_value} and \field{modules.<observations|forecasts|actions|assets>.enabled}. & Event-file columns: \field{event_id}, \field{module}, \field{target_type}, \field{target_id}, \field{target_feature}, \field{start_time_step}, \field{end_time_step}, \field{mode}, with optional \field{value}, \field{std}, \field{min_value}, \field{max_value}, \field{replacement_value} and \field{delay_steps}. Modes depend on the channel: missing/noise/bias/stuck/clip for measurements and forecasts; dropout/noise/bias/stuck/delay/clip for commands; unavailable for equipment. \\
D7 Time resolution and energy units & Top-level \field{seconds_per_time_step}, \field{simulation_start_time_step}, \field{simulation_end_time_step} and \field{episode_time_steps}. Episode selection can additionally use \field{rolling_episode_split} and \field{random_episode_split}. & \field{seconds_per_time_step} is a positive duration such as 3600, 900, 60 or 15 s. Time-series columns declared as power remain in kW; energy columns are in \kwhstep{}. Linked request, deadline and failure fields use integer timestep indices on the same clock. \\
D8 Building and phase power limits & Under \field{buildings.<id>.electrical_service}: \field{mode}, \field{default_split}, \field{limits.total.import_kw}, \field{limits.total.export_kw}, \field{limits.per_phase.<L1|L2|L3>.import_kw} and \field{export_kw}; optional \field{observations} switches expose headroom, violations and phase encoding. Equipment uses \field{attributes.phase_connection}. & \field{mode} is \field{single_phase} or \field{three_phase}; \field{default_split} is \field{balanced}, \field{L1}, \field{L2} or \field{L3} where compatible; \field{phase_connection} is \field{L1}, \field{L2}, \field{L3} or \field{all_phases}. Omitted power limits are unbounded. \\
\bottomrule
\end{longtable}
}

\subsection{Related service and evaluation configuration}

The following settings define EV service, controller inputs, applied actions and result exports.

{\scriptsize
\setlength{\tabcolsep}{3pt}
\begin{longtable}{@{}p{0.23\linewidth}p{0.33\linewidth}p{0.36\linewidth}@{}}
\caption{Related service and evaluation configuration.}\label{tab:related_config_map}\\
\toprule
Item & Main fields or files & Purpose \\
\midrule
\endfirsthead
\toprule
Item & Main fields or files & Purpose \\
\midrule
\endhead
EV service & EV connection and departure schedules, charger power limits, V2G availability, target/minimum state of charge and departure-service fields & Preserves mobility requirements next to grid objectives so a low cost or peak does not hide a missed departure target. \\
Controller inputs and applied actions & \texttt{interface}, structured-input sets, action masks and action feedback & Defines what the controller receives and keeps requested and applied actions separate. \\
Performance indicators and result exports & KPI/export settings, output fields and CSV or Parquet result format & Defines the indicators, trajectories and diagnostics retained for comparison and audit. \\
\bottomrule
\end{longtable}
}

\section{KPI catalogue and equations}
\label{app:kpis}

This appendix summarizes the KPI families in \cref{tab:kpi_families}, with equations where they clarify scope or normalization.  

{\scriptsize
\setlength{\tabcolsep}{3pt}
\begin{longtable}{@{}L{0.13\linewidth}L{0.22\linewidth}L{0.36\linewidth}L{0.12\linewidth}@{}}
\caption{Expanded KPI catalogue for CityLearn v3.}\label{tab:kpi_equations}\\
\toprule
Family & KPI & Definition / rationale & Unit \\
\midrule
\endfirsthead
\toprule
Family & KPI & Definition / rationale & Unit \\
\midrule
\endhead
Energy and power limits & Net electricity consumption & $E^{\mathrm{net}}_{b,t}$ or $E^{\mathrm{net}}_{\mathcal{C},t}$ after equipment operation and local accounting. & kWh/step \\
Energy and power limits & Net-boundary grid import & $\sum_t \max(E^{\mathrm{grid}}_{t},0)$ for signed energy at the stated boundary; distinguish this from summed member imports. & kWh \\
Energy and power limits & Net-boundary grid export & $\sum_t \max(-E^{\mathrm{grid}}_{t},0)$ at the same boundary; distinguish this from summed member exports. & kWh \\
Energy and power limits & Net grid energy & Grid import minus grid export. & kWh \\
Energy and power limits & Average daily peak & Mean of daily maximum net import power. & kW \\
Energy and power limits & Peak import & Maximum net grid import power over the evaluation window. & kW \\
Energy and power limits & Peak export & Maximum net grid export power over the evaluation window. & kW \\
Energy and power limits & Ramping & Sum or mean absolute changes in net load over the evaluation window. & kW or kWh \\
Energy and power limits & Load factor & Average import divided by peak import over the evaluation window. & ratio \\
Energy and power limits & One minus load factor & Complement of load factor when lower-is-better normalization is used. & ratio \\
Energy and power limits & Zero-net-energy ratio & Evaluated-controller net energy relative to the declared reference policy; the exported metric specifies the aggregation scope. & ratio \\
Energy and power limits & Requested-pressure / residual-violation counts & Separate counts of requests beyond configured electrical limits and violations remaining in the applied power state. & count \\
Cost and settlement & Electricity cost & Energy import/export/local settlement multiplied by configured tariffs. & currency \\
Cost and settlement & Local export credit & Locally shared export multiplied by the local price; residual grid export is unremunerated in the described implementation. & currency \\
Cost and settlement & Net energy cost & Import cost minus export/local revenues, using the configured settlement. & currency \\
Emissions & Carbon emissions & $\sum_t E^{\mathrm{grid-import}}_t c_t$, where $c_t$ is carbon intensity. & kgCO$_2$ \\
Emissions & Import carbon intensity & Emissions divided by imported energy. & kgCO$_2$/kWh \\
Inherited thermal/\allowbreak HVAC/\allowbreak comfort & Heat-pump electricity & Electricity consumed by heating/cooling heat-pump devices in scenarios with those devices. & kWh \\
 & Unmet thermal demand & Heating/cooling demand not served, if represented in the selected scenario. & kWh or ratio \\
 & Comfort violation hours & Timesteps or hours outside configured thermal comfort bounds. & time \\
 & Occupant override count & Number of thermostat/comfort overrides, if occupant feedback is active. & count \\
 & HVAC clipping count & Number of heat-pump/HVAC actions clipped by configured limits. & count \\
PV and local energy sharing & PV generation & Total measured or simulated PV energy. & kWh \\
PV and local energy sharing & PV self-consumption & Locally consumed PV divided by PV generation. & ratio \\
PV and local energy sharing & Self-sufficiency & Demand served by local resources divided by total demand. & ratio \\
PV and local energy sharing & Local import & Energy imported by members from local community surplus. & kWh \\
PV and local energy sharing & Local export & Energy supplied by members for local sharing. & kWh \\
PV and local energy sharing & Local exchange & Matched local surplus and demand before residual grid exchange. & kWh \\
PV and local energy sharing & Local sharing ratio & Local exchange divided by eligible surplus or demand according to the exported demand-side or export-side variant. & ratio \\
PV and local energy sharing & Community surplus & Energy remaining after local matching and demand satisfaction. & kWh \\
PV and local energy sharing & Community deficit & Energy not served by local resources before grid import. & kWh \\
PV and local energy sharing & Settlement cost/revenue & Member or community settlement under local and grid prices. & currency \\
EV service & EV connected time-step count & Number of timesteps with EV connected and controllable/observable. & count \\
EV service & EV arrival count & Number of EV arrivals in the evaluation window. & count \\
EV service & EV departure count & Number of EV departures in the evaluation window. & count \\
EV service & Charged energy & Energy delivered to EV batteries. & kWh \\
EV service & Discharged energy / V2G energy & Energy discharged from EVs to building/community/grid. & kWh \\
EV service & Required departure energy & Energy required to reach target SOC at departure. & kWh \\
EV service & Departure success count & Number of departures meeting target SOC within tolerance. & count \\
EV service & Departure success ratio & Successful departures divided by total evaluated departures. & ratio \\
EV service & Departure deficit energy & $\sum_i \max(E^{\mathrm{target}}_i - E^{\mathrm{stored}}_{i,d_i},0)$, where $i$ identifies a departure, $d_i$ is its timestep and the two energy terms are required and stored EV energy. & kWh \\
EV service & Maximum individual departure deficit & Maximum deficit observed for a departing EV. & kWh \\
EV service & Mean departure deficit & Mean deficit across departures. & kWh \\
EV service & Average required charging power & Required energy divided by remaining connection time. & kW \\
EV service & Slack time / urgency & Time remaining after accounting for required charging at feasible power. & time \\
EV service & Charger clipping count & Number of EV actions clipped by charger/battery/electrical constraints. & count \\
Flexible-load service & Requested cycles & Number of flexible-load cycles requested by the scenario. & count \\
Flexible-load service & Started cycles & Number of requested cycles started. & count \\
Flexible-load service & Completed cycles & Number of requested cycles completed before deadline. & count \\
Flexible-load service & Missed cycles & Number of cycles not completed by deadline. & count \\
Flexible-load service & Service level & Completed cycles divided by requested cycles. & ratio \\
Flexible-load service & Cycle energy served & Energy consumed by completed or active cycles. & kWh \\
Flexible-load service & Unserved cycle energy & Requested cycle energy not delivered by deadline. & kWh \\
Flexible-load service & Average start delay & Mean delay relative to earliest feasible start. & time \\
Flexible-load service & Maximum start delay & Maximum delay among served cycles. & time \\
Flexible-load service & Deadline violation count & Number of cycle deadlines missed or violated. & count \\
Demand-response requests & Request count & Number of demand-response requests in the evaluation window. & count \\
Demand-response requests & Active timestep count & Number of timesteps inside active request windows. & count \\
Demand-response requests & Requested total energy & $\sum_{(r,t) \in \mathcal{W}} R_r\Delta t/3600$, where $\mathcal{W}$ contains valid active request--timestep pairs and $\Delta t$ is the timestep in seconds. & kWh \\
Demand-response requests & Delivered total energy & $\sum_{(r,t) \in \mathcal{W}} d_{r,t}\Delta t/3600$, retaining delivery opposite to the requested direction as a negative value. & kWh \\
Demand-response requests & Credited total energy & $\sum_{(r,t) \in \mathcal{W}} c_{r,t}\Delta t/3600$, where $c_{r,t}$ is the non-negative capped credit in \cref{eq:dr_credit_shortfall}. & kWh \\
Demand-response requests & Shortfall total energy & $\sum_{(r,t) \in \mathcal{W}} s_{r,t}\Delta t/3600$, with $s_{r,t}$ defined in \cref{eq:dr_credit_shortfall}. & kWh \\
Demand-response requests & Compliance ratio & Credited energy divided by requested energy over valid request timesteps. & ratio \\
Demand-response requests & Revenue total & Credited energy multiplied by the activation price. & currency \\
Demand-response requests & Penalty total & Shortfall energy multiplied by the penalty price. & currency \\
Demand-response requests & Net revenue total & Revenue minus penalties. & currency \\
Demand-response requests & Invalid baseline timestep count & Number of request timesteps without a valid baseline. & count \\
Demand-response requests & Up/down delivered energy & Signed delivered energy separated by request direction. & kWh \\
Data and equipment failures & Failure count & Number of configured failures activated. & count \\
Data and equipment failures & Changed-measurement count & Number of measurement entries affected by missing, noise, bias, stuck or clipping modes. & count \\
Data and equipment failures & Changed-forecast count & Number of forecast entries affected. & count \\
Data and equipment failures & Missing-measurement count & Number of measurement values set to the missing marker. & count \\
Data and equipment failures & Changed-action count & Number of commands modified by dropout, noise, bias, stuck, delay or clipping modes. & count \\
Data and equipment failures & Lost-command count & Number of commands dropped or replaced. & count \\
Data and equipment failures & Equipment-unavailable count & Number of equipment--timestep pairs marked unavailable. & count \\
Data and equipment failures & Failure summary & Count or weighted count by affected channel and failure type. & mixed \\
Changing community members and assets & Active member count & Number of active REC members per timestep or average over the evaluation window. & count \\
Changing community members and assets & Active asset count & Number of active assets per timestep or average over the evaluation window. & count \\
Changing community members and assets & Change count & Number of member or asset additions/removals. & count \\
Changing community members and assets & Active-period energy & Energy and KPIs computed only over active periods. & mixed \\
Controller inputs and actions & Input table count & Number of structured input tables returned by the interface. & count \\
Controller inputs and actions & Input row count & Number of identified rows across input tables. & count \\
Controller inputs and actions & Relation count & Number of exposed relations, e.g., building--equipment or charger--EV. & count \\
Controller inputs and actions & Action mask count & Number of unavailable/infeasible action entries. & count \\
Controller inputs and actions & Reduced-action count & Number of requested actions reduced by physical or service limits. & count \\
Controller inputs and actions & Feasible action capacity & Available feasible charge/discharge or start capacity exposed to the controller. & mixed \\
Multiple communities & Combined grid import/export & Sum of community import/export. & kWh \\
Multiple communities & Combined cost and emissions & Sum of community cost/emissions. & mixed \\
Multiple communities & Weighted combined ratio & Weighted mean of ratios such as self-consumption or service level. & ratio \\
Multiple communities & Community-level KPI table & Indicators retained for each community before combination. & mixed \\
Multiple communities & Combined request/failure/change totals & Sum of demand-response requests, failures and member/equipment changes across communities. & count \\
\bottomrule
\end{longtable}
}

\par{}

\section{Implementation map}
\label{app:implementation_map}

\begin{longtable}{p{0.18\linewidth}p{0.24\linewidth}p{0.47\linewidth}}
\caption{Implementation areas supporting the scenario and evaluation capabilities.}\label{tab:implementation_map}\\
\toprule
Area & Scope & Role in the paper \\
\midrule
\endfirsthead
\toprule
Area & Scope & Role in the paper \\
\midrule
\endhead
Mobility and flexible services & EVs, chargers and appliance cycles & Represents charging obligations and flexible service windows. \\
Core simulation & Buildings, storage, PV, control loop, established KPIs and datasets & Provides the simulation basis for the REC requirements in \cref{sec:v2_to_requirements}. \\
Software distribution & Official CityLearn repository and versioned releases & Distributes the simulator and its public examples. \\
Control loop and physical corrections & Terminal states, action parsing, SOC, storage and high-frequency timestep handling & Supports explicit scenarios and comparable controller studies. \\
Performance indicators and exports & Runtime, loading, export and KPI separation & Supports auditable KPI calculation and result inspection. \\
Structured inputs and changing members/assets & Identified controller inputs and changing active periods & Supports structured control and changing community members and assets. \\
Time resolution and Parquet & Power/energy units, measured PV and high-frequency data loading & Supports measured data and large datasets. \\
Flexible-load schedules and deadlines & Cycle profiles and allowed start windows & Supports flexible loads beyond washing machines. \\
Demand-response requests & File-based requests and settlement indicators & Supports demand-response evaluation. \\
Multiple communities & Synchronized runs and weighted combination & Supports several independent RECs with individual and combined KPIs. \\
Data and equipment failures & Changes to measurements, forecasts, commands and availability & Supports repeatable failure testing. \\
\bottomrule
\end{longtable}

\end{document}